\documentclass[10pt,journal,doublecolumn]{IEEEtran}
\usepackage{amsthm}
\usepackage{cite}
\usepackage{amssymb,amsmath}
\usepackage{times}
\usepackage{latexsym}
\usepackage{bm}
\usepackage{stfloats}
\usepackage{algorithm}
\usepackage{algpseudocode}
\usepackage{cases}
\usepackage{array}
\usepackage{xcolor}
\usepackage{setspace}
\usepackage{fancyhdr}
\usepackage[small]{caption}
\usepackage{soul}
\usepackage{subfig}
\usepackage{graphicx}
\usepackage{fixltx2e}
\usepackage{multicol}
\usepackage{latexsym}
\usepackage{bm}
\usepackage{enumerate}  

\newcommand{\Nrx}{N_{\rm rx}}
\newcommand{\Ntx}{N_{\rm tx}}
\newcommand{\Nmp}{N_{\rm mp}}
\newcommand{\Nsc}{N_{\rm sc}}
\newcommand{\Nnz}{N_{\rm nz}}
\newcommand{\Nss}{N_{\rm ss}}
\newcommand{\Nltf}{N_{\rm ltf}}
\newcommand{\Ncp}{N_{\rm cp}}
\newcommand{\Ngrd}{N_{\rm g}}
\newcommand{\Ntot}{N_{\rm t}}
\newcommand{\Npkt}{N_{\rm p}}
\newcommand{\Npart}{N_{\rm b}}

\newcommand{\ChanMx}{{\bf H}}
\newcommand{\ChanElem}{\mathfrak{h}}

\newcommand{\CirVec}{{\bf h}}
\newcommand{\CirElem}{h}
\newcommand{\pdp}{\mathcal{P}}
\newcommand{\CSIMx}{{\bf CSI}}
\newcommand{\CSIVec}{{\bf csi}}
\newcommand{\CSIElem}{{\rm csi}}
\newcommand{\CDDa}{{\bf \Phi_a}}
\newcommand{\CDDb}{{\bf \Phi_b}}
\newcommand{\SMM}{{\bf Q}}
\newcommand{\RecMx}{{\bf Y}}
\newcommand{\OrthMx}{{\bf P}}
\newcommand{\OrthElem}{p}
\newcommand{\scidx}{k}

\newcommand{\timeDisc}{n}
\newcommand{\delDisc}{m}

\newcommand{\bw}{B}
\newcommand{\txidx}{j_{\rm tx}}
\newcommand{\rxidx}{j_{\rm rx}}
\newcommand{\mpidx}{j_{\rm mp}}
\newcommand{\ltfidx}{j_{\rm ltf}}
\newcommand{\ssidx}{j_{\rm ss}}
\newcommand{\pktidx}{j_{\rm p}}
\newcommand{\partidx}{j_{\rm b}}
\newcommand{\freeidx}{r}

\newcommand{\identityMx}{\bf I}
\newcommand{\antidiagMx}{\bf J}

\newcommand{\ReceElem}{y}
\newcommand{\TransElem}{x}
\newcommand{\TransVec}{{\bf x}}
\newcommand{\CDDaElem}{\phi_a}
\newcommand{\CDDbElem}{\phi_b}
\newcommand{\CDDaCoeff}{{\delta_a}}
\newcommand{\CDDbCoeff}{{\delta_b}}

\newcommand{\SMMElem}{q}
\newcommand{\SampInt}{T_{\rm s}}

\newcommand{\AttCoeff}{\beta}
\newcommand{\CompAttCoeff}{\gamma}
\newcommand{\DelCoeff}{\tau}
\newcommand{\freq}{f}
\newcommand{\ScSpacing}{\Delta \freq}

\newcommand{\WinFreq}{W}
\newcommand{\WinTime}{w}
\newcommand{\PsiOne}{\psi}
\newcommand{\PsiMx}{\Psi}

\newcommand{\ChanPhLin}{\epsilon^{\rm ch}}
\newcommand{\Preadvancement}{\epsilon^{\rm pre}}

\newcommand{\CFOoff}{\Delta_c}
\newcommand{\CPOoff}{\phi_c}
\newcommand{\CFOCoeff}{\zeta_{\rm CFO}}
\newcommand{\SFOCoeff}{\zeta_{\rm SFO}}
\newcommand{\STOcount}{N_{\rm STO}}

\newcommand{\CFOCalibFunc}{g_2}
\newcommand{\SFOCalibFunc}{g_1}

\newcommand{\SteeringMx}{{\bf A}}
\newcommand{\SourceMx}{{\bf \gamma}}

\newcommand{\NoiseSamp}{\mathsf{n}}
\newcommand{\NoiseVec}{{\bf n}}
\newcommand{\NoiseMx}{{\bf N}}

\newcommand{\SteeringVec}{{\bf a}}
\newcommand{\NoisePower}{\sigma}
\newcommand{\CovMx}{{\bf R}}

\newcommand{\EigValMx}{\Lambda}
\newcommand{\NoiseEigVec}{{\bf e}}
\newcommand{\NoiseEigMx}{{\bf E}_n}

\newcommand{\EigVecMx}{{\bf E}}

\newcommand{\distEst}{\hat{d}}
\newcommand{\distTruth}{d^{\rm truth}}

\algnewcommand{\algorithmicgoto}{\textbf{go to}}%
\algnewcommand{\Goto}[1]{\algorithmicgoto~\ref{#1}}%

\begin{document}

\title{Decimeter Ranging with Channel State Information}
\author{Navid~Tadayon,~\IEEEmembership{Member,~IEEE}, Muhammed~T.~Rahman,  Shuo~Han, Shahrokh~Valaee,~~\IEEEmembership{Senior Member,~IEEE}, and Wei~Yu,~~\IEEEmembership{Fellow,~IEEE}
\thanks{This research is supported by the Natural Sciences and Engineering Research Council (NSERC).}
\thanks{Aurthors are affiliated with the University of Toronto (UofT), Toronto, Canada; Email: tadayon@utoronto.ca, \{mt.rahman, shuo.han\}@mail.utoronto.ca, and \{valaee, weiyu\}@ece.utoronto.ca.}
}
\maketitle
\begin{abstract}
This paper aims at the problem of time-of-flight (ToF) estimation using channel state information (CSI) obtainable from commercialized MIMO-OFDM WLAN receivers. It was often claimed that the CSI phase is contaminated with errors of known and unknown natures rendering ToF-based positioning difficult. To search for an answer, we take a bottom-up approach by first understanding CSI, its constituent building blocks, and the sources of error that contaminate it. We then model these effects mathematically. The correctness of these models is corroborated based on the CSI collected in extensive measurement campaign including radiated, conducted and chamber tests. Knowing the nature of contaminations in CSI phase and amplitude, we proceed with introducing pre-processing methods to clean CSI from those errors and make it usable for range estimation. To check the validity of proposed algorithms, the MUSIC super-resolution algorithm is applied to post-processed CSI to perform range estimates. Results substantiate that median accuracy of $0.6$m, $0.8$m, and $0.9$m is achievable in highly multipath line-of-sight environment where transmitter and receiver are $5$m, $10$m, and $15$m apart.
\end{abstract}



\section{Introduction}\label{sec.intro}
\IEEEPARstart{O}{ne}  of the fundamental challenges of today's networks is precise estimation of indoor users' locations. The location of a user is a source of information that can be leveraged to unlock huge technological, social, and business potentials. This is in particular the case for indoor environment, where the signal of the global navigation satellite system (GNSS) is unavailable. 

Due to its pervasive deployment and cost-effective nature, positioning using wireless local area networks (WLANs) signals has been at the focus of research for almost a decade. In fact, experimental works have proven that WiFi signals can be used to obtain excellent location accuracy even in harsh multipath environments \cite{ferris2007, kotaru2015, xiong2013, sen2013, vasisht2016, mariakakis2014, yang2015, chintalapudi2010, bahl2000radar, youssef2005horus,azizyan2009surroundsense}. For a comprehensive survey on the success of WiFi in localizing indoor users refer to \cite{xiao2016survey,yang2013rssi}. This has been a significant advancement as, until recently, ultra-wide-band (UWB) radio was deemed as the only viable solution to get accurate location information \cite{Gezici2005}. 

Indoor positioning using WiFi began with power-based ranging using received signal strength (RSS) \cite{chintalapudi2010, bahl2000radar, youssef2005horus, azizyan2009surroundsense, SorourMay2015, FengDec2012}. 
Unfortunately, accurate range estimation with RSS is impossible because: \textbf{(i)} time-domain OFDM signals are highly fluctuating \textbf{(ii)} the amplitude of a signal is directly affected by small-scale fading \textbf{(iii)} signal amplification at the receiver is controlled by the automatic gain controller (AGC) whose behaviour dynamically varies with channel conditions. 

This paper is motivated by the availability of channel-state information (CSI) from Intel \cite{halperin2011tool} and Atheros \cite{xie2015precise} WiFi chipsets that have enabled CSI to be used for positioning. CSI is a more stable and informative representation of the wireless channel (compared to RSS) between two communicating end-points. Therefore, it can be used to perform range (time-based or power-based) and angle-of-arrival estimation. When it comes down to implementation, while CSI-based localization with AoA achieved promising outcomes \cite{kotaru2015, xiong2013, sen2013,Rahman2018}, using CSI to estimate time-of-flight (ToF) measurement has either not been pursued or led to inconsistent results \cite{kotaru2015}. {{To our knowledge, the studies that do consider phase-based ranging all use software-defined radio (SDR), an open-source and fine-tuned platform that is expensive to acquire and so is unscalable. On the other hand, our work is based on using commercial off-the-shelf MIMO-OFDM network interface cards (NIC), which are used in laptops and computers, to estimate the range from phase of the CSI.}} As ToF measurement is crucial to ranging, and subsequently positioning, this raises the question, ``What makes ToF estimation using CSI a challenging task?" This paper aims to find an answer to this question. Our goal is multifaceted: First, we aim to discuss some of the often neglected practical issues about CSI and ToF estimation using the CSI. In that vein, we dissect CSI that is obtainable from WiFi chipsets to understand its constituent building blocks, different forms it takes, and the sources of error that contaminate it. We proceed with introducing pre-processing methods to clean CSI from those errors and make it usable for ToF estimation. We then apply the classic super-resolution spectral MUSIC algorithm to the post-processed CSI to obtain accurate and stable range estimates. {{To our knowledge, this is an achievement that has never been accomplished before. }}

 The inherent appeal of MUSIC algorithm is due to the fact that estimator's resolvability power is not only determined by the signal bandwidth but also the total signal-to-noise ratio (SNR). {{More importantly, MUSIC is an efficient and consistent estimator when certain criteria are met.}}
 
  In doing so, different ideas are examined, including covariance hardening methods, such as spectral smoothing and forward-backward smoothing, and decision fusion algorithms. We demonstrate that decimetre ranging with only $20$MHz of spectrum is possible if CSI is properly post-processed and range estimates are intuitively combined.

\begin{figure*}[t!]
\centering
\subfloat[Macroscopic view of the system.]{\label{fig0a}\includegraphics[scale=0.45]{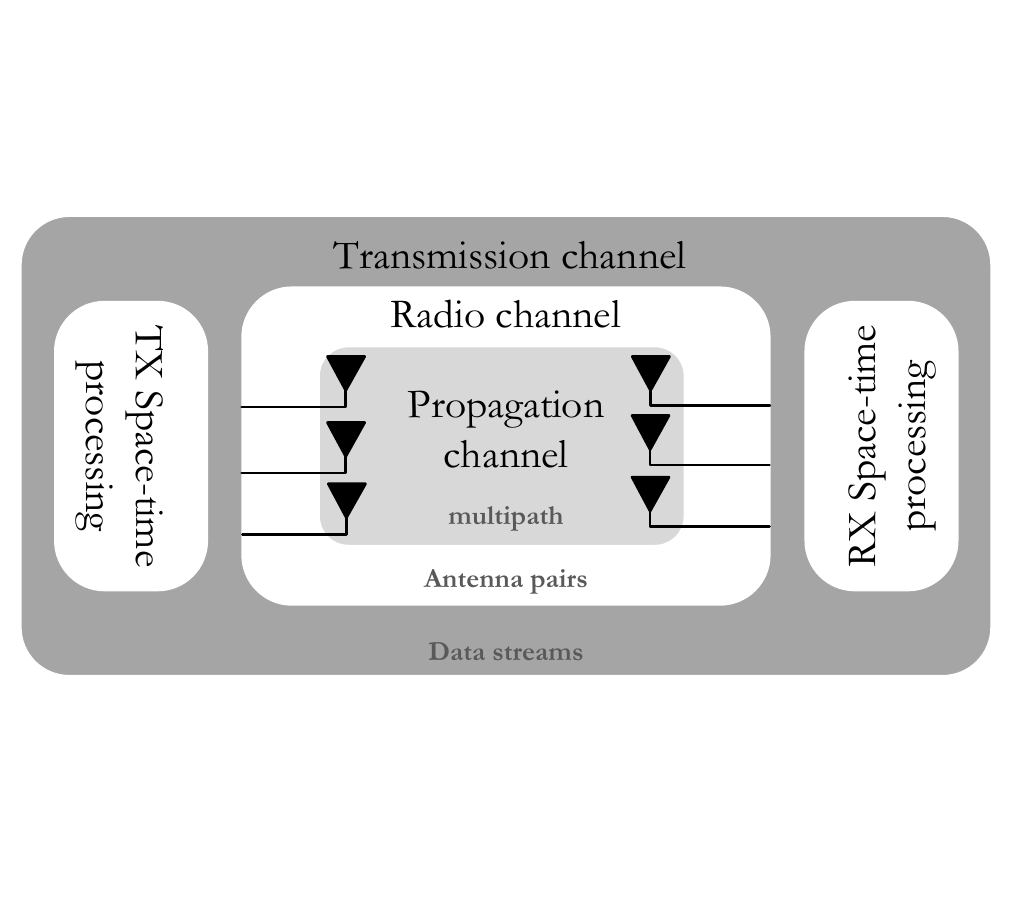}} \hspace{1cm}
\subfloat[MIMO-OFDM transceiver architecture.]{\label{fig0b}\includegraphics[scale=0.5]{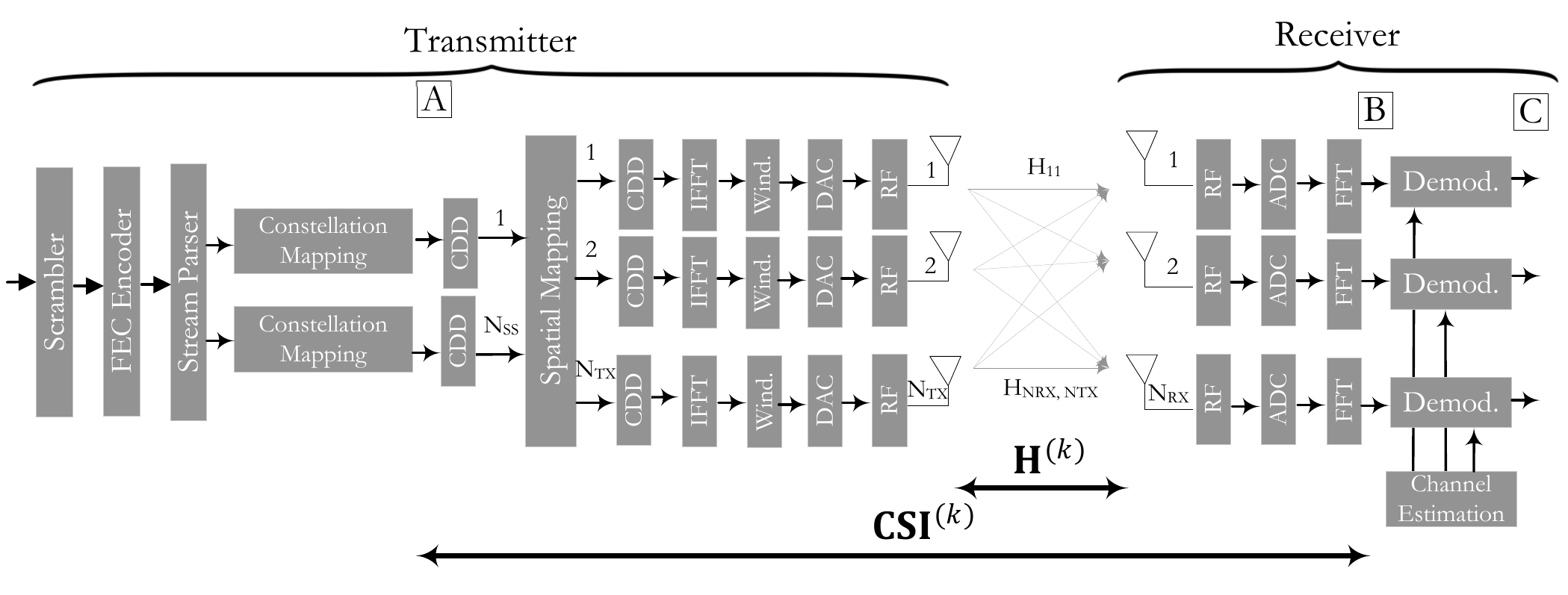}}
\caption{{{(left) Depiction of different channels observed at different points along the tranceiver chain (right) End-to-end MIMO-OFDM transceiver architecture. Channel from point A to C is the equivalent transmission channel.}}}
\label{fig0}
\end{figure*}

{\bf Problem Statement: } A holistic view of the problem addressed in this paper is presented in Fig. \ref{fig0a} where the link between a transmitter and receiver is shown: {{Whereas  coherent decoding of data symbols in communications systems requires the knowledge of end-to-end degradation imposed between a transmitter's baseband (BB) and receiver's BB (named \textit{transmission} channel), location estimation hinges on the knowledge of the channel immediately between the two antennas (named \textit{propagation} channel). Not only these two channels are not the same, but quantifying one from the other is a non-trivial task.}} The difference between transmission channel MX $\CSIMx$ and propagation channel $\ChanMx$ arises because of \textit{(i)}  lack of synchronization between transmitter/receiver in passband (PB) and \textit{(ii)}  deterministic signal processing operations in transmitter's BB. In the latter case, cyclic delay diversity (CDD), spatial mapping matrix (SMM), and time-windowing, whose effects are generally incorporated into the CSI matrix, make the receiver believe that the transmitter is several tens of meters away and that the channel is more reflective than it really is. 

{\bf Contribution: } In tackling the aforementioned problem, this paper's contributions are as follows:
\begin{itemize}
\item To dissect different deterministic and random phenomena happening in the transmitter and receiver hardware causing $\ChanMx$ and $\CSIMx$ to be different. 

\item To establish the right model for CSI and its relation with the channel matrix.

\item To develop pre-processing techniques to eliminate random phases introduced by the insufficiency of synchronization between the transmitter and the receiver.  

\item To obtain accurate range estimates by applying super-resolution algorithm to the calibrated CSI.
\end{itemize}

{\bf Organization: } This paper is organized as follows: In Section \ref{Background}, we go over the basics of multiple-input multiple-output orthogonal frequency division multiplexing (MIMO-OFDM) WLAN systems, including their transceiver architecture, channel-sounding, etc. In Section \ref{CSIchallenges} we show why ToF estimation with CSI is a challenging task, and explain different random and deterministic sources of error contributing to this problem. With the knowledge gained, we tackle the problem of cleaning and calibrating CSI in Section \ref{CSIcalib}. Finally, in Section \ref{TOAestimation}, we introduce ideas to obtain more accurate range estimates from the post-processed CSI.

{\bf Notation: } The following notation is adopted throughout this paper:  $a$ (lowercase/regular)$\rightarrow$ a scalar, ${\bf a}$ (lowercase/boldface)$\rightarrow$ a vector, ${\bf A}$ (uppercase/boldface)$\rightarrow$ a matrix. For matrix ${\bf A}$, ${a}_{r,q}$ is its $(r,q)$th element, ${\bf A}^{\rm T}$ is its transpose, ${\bf A}^{\rm *}$ is its conjugate, and ${\bf A}^{\rm H}$ is its Hermitian.

%
%

\section{Background} \label{Background}

\subsection{Channel State Information (CSI)}
Without properly compensating for the propagation and asynchronization effects, the receiver has no way of detecting what was transmitted. To that end, and through a mechanism named \textit{channel sounding}, the receiver obtains an estimate of wireless channel. This is accomplished by sending a training sequence that is known to both transmitter and receiver. For a wideband MIMO-OFDM system, the estimate of the channel is a collection of complex matrices one for each OFDM subcarrier. It is such information that is universally known as channel state information (CSI). Once CSI is known, it is used by the equalizer in order to cancel out any deterioration (e.g. phase shift, attenuation, etc.) that was imposed on the transmitted data. 
For packet-based MIMO-OFDM IEEE802.11(n) systems, training sequences, namely high throughput long training fields (HT-LTF), are sent in the preamble, which is instantly used by the receiver to derive CSI. 

\subsection{WLAN Transceiver Architecture}
Fig. \ref{fig0b} shows the general structure of the MIMO-OFDM WLAN transmitter. An encoded high-rate bit stream is fed to the stream parser to create $\Nss$ spatial streams. These spatial streams are modulated using constellation mappers (e.g. QAM) to create stream of symbols. As explained before, the transmitter may only send $\Nss\leq {\rm rank}(\ChanMx)$ parallel streams, where $\ChanMx$ is the true channel matrix, and violating this rule would result in loss of data. {{Note that ${\rm rank}(\ChanMx)\leq \min(\Ntx, \Nrx)$ (with the equality holding when the channel is rich scattering), where $\Ntx$, $\Nrx$ are the number of transmit and receive antennas, respectively.}}

Next, spatial streams are cyclically shifted through a mechanism named cyclic delay diversity (CDD) to create extra frequency diversity and make sure no unintended beamforming takes place when sending common information (e.g. headers) from all transmit antennas. 

The spatial mapping maps fewer number of spatial streams to larger number of transmit antennas \cite{perahia2013next}. This is especially crucial in situations where lower number of streams is to be carried by larger number of transmit chains. 
The existence of CDD and spatial mapping matrix are among the main reasons to render one-way measurements of time-of-flight (ToF) for ranging difficult. Moving forward, a second CDD layer is applied to each transmit chain and frequency domain samples are fed to inverse fast Fourier transform (IFFT) to create time-domain samples. These samples are then simultaneously sent from all transmit chains. 

Referring to Fig. \ref{fig0}, the receiver output ${\bf \ReceElem}$ at point ``B", is related to transmitter input ${\TransVec}$ at point ``A" through the following matrix equation:
\begin{equation} \label{input_ouput_eq}
\begin{split}
{\bf \ReceElem}^{(\scidx)}=&\left[\ChanMx^{(\scidx)}\right]_{ \Nrx \times \Ntx} \cdot \left[\CDDa^{(\scidx)}\right]_{\Ntx \times \Ntx} \cdot 
\\
&
\left[\SMM^{(\scidx)}\right]_{\Ntx \times \Nss} \cdot \left[\CDDb^{(\scidx)}\right]_{\Nss \times \Nss} \cdot {\bf \TransElem}^{(\scidx)} 
\end{split}
\end{equation}
where $\ChanMx^{(\scidx)}$, $\SMM^{(\scidx)}$, $\CDDa^{(\scidx)}$ and $\CDDb^{(\scidx)}$ are, respectively, the channel matrix, the spatial mapping matrix, and the first, and the second CDD matrices at the $\scidx$th subcarriers, $\scidx=1,...,\Nnz$, where $\Nnz$ is the number of (non-zero) subcarriers within the band of interest out of the total of $\Nsc$ subcarriers (e.g. $\Nnz=56$ and $\Nsc=64$ for $B=20$MHz in IEEE 802.11n systems). More details on the composition of $\SMM^{(\scidx)}$,$\CDDa^{(\scidx)}$, and $\CDDb^{(\scidx)}$ are provided in the next sub-section. 

\subsubsection{Cyclic Delay Diversity (CDD)}
Despite that the payload part of a packet is destined only to a given destination, the packet preamble is meant to be heard/decoded by everyone. To ensure that the header is received by all, and to avoid inadvertent  beamforming across the antennas, CDD is used \cite{van2006802}. This is achieved by sending the same header OFDM symbols over different antennas while cyclically shifting them so that \textit{(i)} all RF chains are utilized, thus, longer communication range is obtained \textit{(ii)} no unintended beamforming is experienced. The effect of CDD on transmitting common header information changes the multipath nature of the channel as seen by the receiver.
To simplify the transceiver architecture, CDD is always applied no matter which portion of packet is being sent, header or payload. The choice of CDD is implementation dependent. We observe that at times, even the same access point (AP) will use different CDD values for the same number of streams. Nonetheless, the standard \cite{IEEE80211N} puts forth some recommendations. Ranging with the raw CSI obtained from the NIC (without accounting for CDDs) may give rise to an accuracy that is off by several tens of meters.\footnote{For example, for a 4x4 MIMO system, CDD values $0, -400, -200,-600$ns are suggested. For WLAN systems operating on sampling rate $\SampInt=1/\bw=50$ns, where $\bw=20$MHz, these CDDs are equivalent to delays equivalent to {{$0,8,4,12$}}  samples.}

\subsubsection{Spatial Mapping Matrix (SMM)}

The spatial mapping operation is the most crucial component of MIMO-OFDM systems assuming tasks such as transmit beamforming, spatial multiplexing, spatial diversity, and so on. This is often implemented through linear matrix operation $\SMM^{(\scidx)}$ as shown in \eqref{input_ouput_eq} and is an implementation-dependent matter. If $\Nss=\Ntx$, often direct mapping takes place, i.e. $\SMM^{(\scidx)}=\identityMx$, where $\identityMx$ is the identity matrix. However, when $\Nss<\Ntx$, indirect mapping may be adopted \cite{perahia2013next}. In the latter case, the effect of SMM is similar to having more echoes than those added by the propagation environment. For this reason, imposition of SMM has similar effect as having {\it virtual} echoes.

\subsection{Channel Sounding}
Channel sounding is the mechanism of obtaining CSI at the receiver. This is done by transmitting known HT-LTF sequences. {{HT-LTF sent over $\ssidx$th stream is a unique sequence}} ${\bf \TransElem}_{\ssidx}=(\TransElem_{\ssidx}^{(\scidx)}, \scidx=1\cdots \Nsc)$  where $\TransElem_{\ssidx}^{(\scidx)} \in \{-1,1\}$. To probe a single dimension of the multi-dimensional (MIMO) channel, one ${\bf \TransElem}_{\ssidx}$ is sent on each spatial stream, for the total of $\Nss$ stream. That means that vector ${\bf \TransElem}^{(\scidx)}=(\TransElem_{\ssidx}^{(\scidx)}, \ssidx=1\cdots \Nss)$ is fed to all the $\Nss$ streams simultaneously to be transmitted over the $\scidx$th subcarrier in order to estimate MIMO channel matrix on the $\scidx$th subcarrier frequency. Let's denote ${\bf \hat{x}}=({\bf \TransElem}_{\ssidx}, {\ssidx}=1\cdots\Nss)$.
To probe all the dimensions of the MIMO channel, not one but several ${\bf X}=({\bf \hat{x}}_{\ltfidx}, {\ltfidx}=1\cdots\Nltf)$ are transmitted in the preamble (in sequence) where, $\Nltf \geq \Nss$. In other words, $\Nltf\times \Nss \times \Nsc$ {{two-state training symbols}} $\TransElem_{\ssidx}^{(\scidx)}$ will have to be sent to learn $\Nrx \times \Nss \times \Nsc$ complex coefficients of the MIMO channel \cite{perahia2013next}. Subsequently, a matrix $\RecMx^{(\scidx)}$ is received for the $\scidx$th HT-LTF symbol on $\Nrx$ received antennas. 
%

\section{Challenges of Ranging with CSI} \label{CSIchallenges}
  In general, ToF estimation based on CSI suffers from several deep-rooted issues some of which have not been discussed in the literature. These issues are pointed out next and dealt with in detail later on. 
\paragraph{\bf Bandwidth Limitation} Range estimation has been traditionally done through derivation of the channel impulse response (CIR) for each tx/rx pair and hunting CIR's first and strongest peak. This simple approach has been effective in ranging with UWB radio and been lately pursued in the WiFi-based indoor localization literature \cite{xie2015precise, wu2013csi, yang2013rssi}. Without delving into derivation details, CIR is obtained by taking the IFFT of the samples of the channel-frequency response (CFR), i.e. CSI metric, while accounting for the fact that no CSI is collected on $\scidx=0$ (i.e. zero subcarrier)\footnote{Transmitting data on OFDM's center frequency would result in loss of information due to strong DC current at BB.} and is given by
\begin{equation} \label{CIRformula}
\begin{split}
\CirElem^{(\delDisc)}= \sum_{\mpidx=1}^{\Nmp}{\Gamma_{\mpidx}\bigg(\frac{\sin\big(\frac{\pi(\Nsc+1)}{\Nsc}(\kappa_{\mpidx}-\delDisc)\big)}{\sin\big(\frac{\pi}{\Nsc}(\kappa_{\mpidx}-\delDisc)\big)}-1\bigg)}
\end{split}
\end{equation}
where $\delDisc$ is the time (delay) domain index and
\begin{equation} \nonumber
\begin{split}
\Gamma_{\mpidx} = \frac{1}{\Nsc} \AttCoeff_{\mpidx}e^{-2\pi i \freq_0 \DelCoeff_{\mpidx}} e^{\frac{\pi i (\Nsc-1)}{\Nsc}\delDisc} \mbox{\;\;\; and \;\;\;}
 \kappa_{\mpidx} =\Nsc\Delta\freq\DelCoeff_{\mpidx}
\end{split}
\end{equation}
where $\Nmp$, $\DelCoeff_{\mpidx}$, $\AttCoeff_{\mpidx}$, $\ScSpacing$, $\freq_0$ are the number of multipath arrivals, delay and  attenuation on $\mpidx$th path, subcarrier-spacing, and central frequency, respectively. This power-delay-profile (PDP) peaks at discrete samples {{$\delDisc = \delDisc_{\rm peak}=\lfloor \kappa_{\mpidx} \rceil$}} only if \textit{(i)} $\mpidx$th arrival has enough strength $|\Gamma_{\mpidx}|$ \textit{(ii)} close-by arrivals are not within each other's Rayleigh resolution limit, i.e. $|\DelCoeff_{\mpidx}-\DelCoeff_{\mpidx^\prime}|>1/(\Nsc\ScSpacing)$. For WiFi systems with sampling rate  $20$Mega sample/s (Msps) (for a $\bw=20$MHz channel), the electromagnetic wave travels extra $15$m between two consecutive samples. Such low sampling rate makes resolving closely-spaced multipath reflections (as needed for indoor positioning) based on CIR theoretically impossible.
\paragraph{\bf CSI Phase Contamination} The phase in the CSI matrix is contaminated with terms triggered by the imperfect synchronization between the transmitter and receiver in analog/digital domains. Dubbed by the names symbol timing offset (STO), sampling frequency offset (SFO), carrier frequency offset (CFO), and carrier phase offset (CPO), these frequency and time synchronization errors are extremely volatile in nature \cite{speth1999optimum}. 
\paragraph{\bf CSI Amplitude Contamination} The amplitude of the CSI is highly distorted by three phenomena: (a) unpredictable changes in AGC gain, (b) I/Q imbalance, and (c) the mixed effect of cyclic-prefix removal/guard-band insertion/windowing operation on time-domain CSI samples. 
\paragraph{\bf CDD Phase Shift} The CDD included in the CSI matrix appears as an additive phase in the CSI matrix. CDD can potentially degrade the ranging accuracy using CSI by several tens of meters. This is particularly the case when $\Nss<\Ntx$ \cite{perahia2013next, van2006802}.

\paragraph{\bf Artificial Multipath} The multiplexing operation $\SMM^{(\scidx)}$ performed on input streams causes the received samples to look as if they were transmitted on a fading channel with many more reflections \cite{IEEE80211N, perahia2013next}.

\subsection{Impact of OFDM Baseband Operations}


\subsubsection{SMM and CDD}
Accounting for the SMM and CDD operations at the transmitter, the entire sounding mechanism can be described by \eqref{eq.3}, at the top of the page, where $\OrthMx$ (the rightmost matrix) is called the orthogonal mapping matrix.
\begin{figure*}[!btp]
\begin{equation}\label{eq.3} \small 
\begin{aligned}
{\begin{bmatrix}
 \ReceElem_{1,1}^{(\scidx)}&\cdots   &\ReceElem_{1,\Nltf}^{(\scidx)} \\ 
                \vdots & \ddots  &\vdots  \\ 
 \ReceElem_{\Nrx,1}^{(\scidx)}&\cdots&\ReceElem_{\Nrx,\Nltf}^{(\scidx)} 
\end{bmatrix}}
=&{\begin{bmatrix}
 \ChanElem_{1,1}^{(\scidx)}&\cdots   &\ChanElem_{1,\Ntx}^{(\scidx)} \\ 
 \vdots & \ddots  &\vdots  \\ 
 \ChanElem_{\Nrx,1}^{(\scidx)}&\cdots   &\ChanElem_{\Nrx,\Ntx}^{(\scidx)} 
\end{bmatrix}}
{\begin{bmatrix}
 {{\CDDaElem}_{1,1}^{(\scidx)}}&\cdots&0 \\ 
 \vdots & \ddots  &\vdots  \\ 
 0&\cdots   & {{\CDDaElem}_{\Ntx,\Ntx}^{(\scidx)}} 
\end{bmatrix}} \times \\
& {\begin{bmatrix}
 \SMMElem_{1,1}^{(\scidx)}&\cdots   &\SMMElem_{1,\Nss}^{(\scidx)} \\ 
        \vdots & \ddots  &\vdots  \\ 
\SMMElem_{\Ntx,1}^{(\scidx)}&\cdots   &\SMMElem_{\Ntx,\Nss}^{(\scidx)}
\end{bmatrix}} 
{\begin{bmatrix}
{{\CDDbElem}_{1,1}^{(\scidx)}}&\cdots&0 \\ 
 \vdots & \ddots  &\vdots  \\ 
 0&\cdots   & {{\CDDbElem}_{\Nss,\Nss}^{(\scidx)}} 
\end{bmatrix}}
 {\begin{bmatrix}
 \TransElem_{\scidx}\OrthElem_{1,1}&\cdots   &\TransElem_{\scidx}\OrthElem_{1,\Nltf} \\ 
        \vdots & \ddots  &\vdots  \\ 
 \TransElem_{\scidx}\OrthElem_{\Nss,1}&\cdots   &\TransElem_{\scidx}\OrthElem_{\Nss,\Nltf}
\end{bmatrix}}
+\NoiseMx_{\scidx}
\end{aligned}
\end{equation}
\rule{\textwidth}{0.5pt}
\end{figure*}
The CSI matrix is calculated  as $\CSIMx^{(\scidx)}=\RecMx^{(\scidx)}\OrthMx^{-1}$, for each subcarrier. In \eqref{eq.3}, ${\CDDa}$ and ${\CDDb}$ are the cyclic shift (diagonal) matrices before and after spatial mapping, which is denoted by $\SMM$, a linear matrix, as shown in the transceiver architecture of Fig. \ref{fig0} and $\NoiseMx_{\scidx}$ is the noise matrix. Because $\CDDa$, $\CDDb$, $\SMM$ are implementation-dependent quantities, estimating matrix $\ChanMx^{(\scidx)}$ at the receiver from observations $\CSIMx^{(\scidx)}$ is challenging. However, the receiver does not require to extract the channel matrix $\ChanMx^{(\scidx)}$ to decode data points; so long as $\CDDa, \CDDb, \SMM$  are applied to both training sequences and payload (which is indeed the case), the receiver can view $\hat{\ChanMx}^{(\scidx)}=\ChanMx^{(\scidx)}\CDDa^{(\scidx)}\SMM^{(\scidx)}\CDDb^{(\scidx)}$ as an end-to-end channel. 
Elaborating on \eqref{eq.3}, and given that the receiver removes the orthogonal mapping matrix $\OrthMx$, the $(\rxidx,\ssidx)$ element of the CSI matrix is given by
\begin{equation}\label{eq.4}
\begin{split}
\CSIElem_{\rxidx,\ssidx}^{(\scidx)}= \sum_{\txidx=1}^{\Ntx}\ChanElem_{\rxidx,\txidx}^{(\scidx)}{\CDDaElem}_{\txidx,\txidx}^{(\scidx)}\SMMElem_{\txidx,\ssidx}^{(\scidx)}{\CDDbElem}_{\ssidx,\ssidx}^{(\scidx)}+\NoiseSamp_{\scidx}
\end{split}
\end{equation}
where $\rxidx$, $\txidx$, $\ssidx$ represent receive antenna, transmit antennas, and spatial stream indices, respectively. From \eqref{eq.4}, the information on the ToF of the line-of-sight (LoS) path is concealed in $\ChanElem_{\rxidx,\txidx}^{(\scidx)}$ which is given by
\begin{equation}\label{eq.4.1}
\begin{split}
\ChanElem_{\rxidx,\txidx}^{(\scidx)}=\sum_{\mpidx=1}^{\Nmp}{\AttCoeff_{\mpidx}^{\rxidx,\txidx}e^{-2\pi i \freq_{\scidx}\DelCoeff_{\mpidx}^{\rxidx,\txidx}}}
\end{split}
\end{equation}
where $\AttCoeff_{\mpidx}^{\rxidx,\txidx}$ and $\DelCoeff_{\mpidx}^{\rxidx,\txidx}$ are the attenuation and time delay of the $\mpidx$th path between $\rxidx$th receive and $\txidx$th transmit antennas, respectively. Also, $\Nmp$ is the number of multipath components and $\freq_{\scidx}=\freq_0+\scidx \ScSpacing$ is the $\scidx$th subcarrier's frequency with $\ScSpacing$ and $\freq_0$ being the subcarrier spacing and the center frequency, respectively. 
\begin{figure}[t!]
\centering
\subfloat[Chamber, $\Nss=2$.]{\label{PDPChamb.a}\includegraphics[scale=0.4]{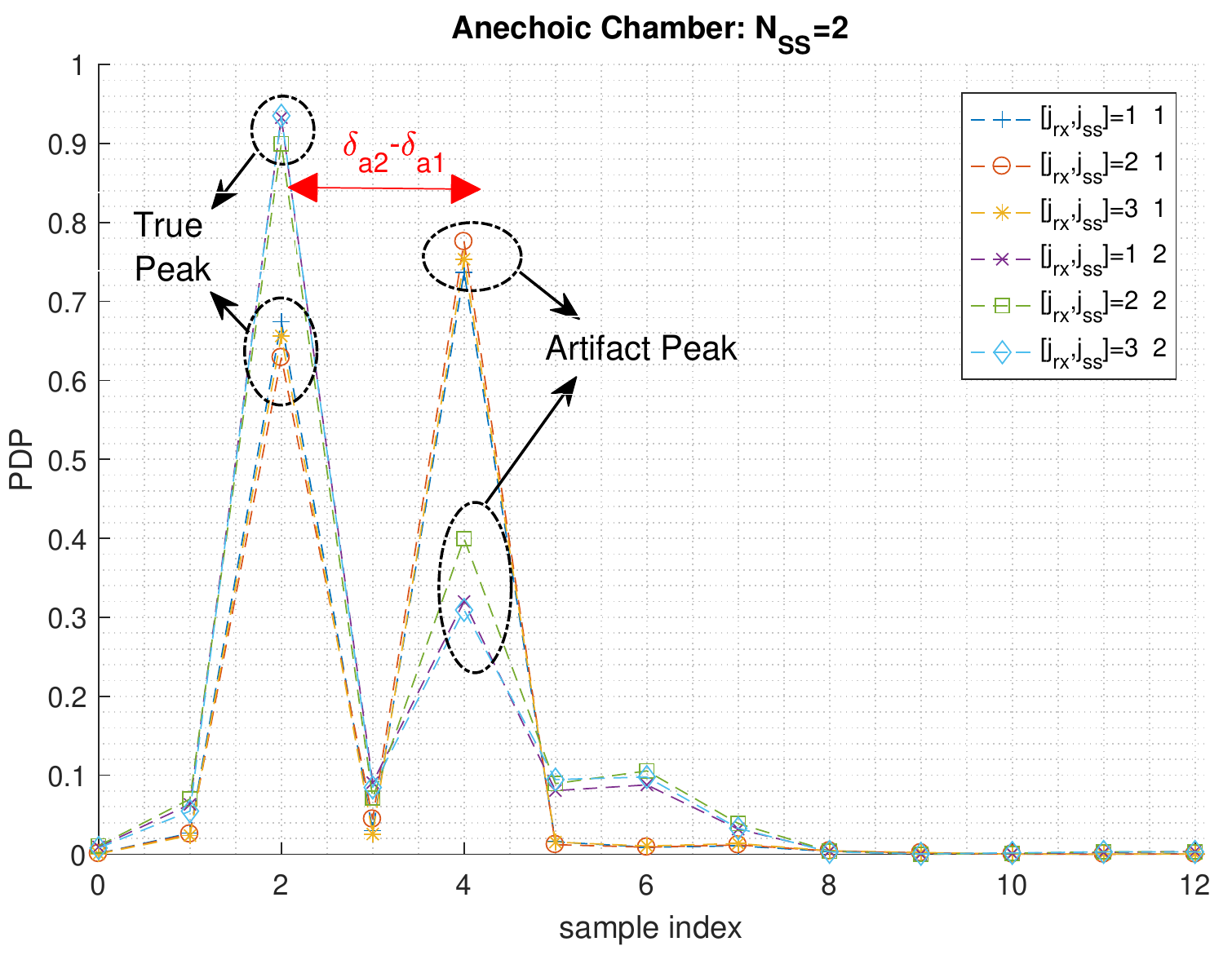}} \\
\subfloat[Chamber, $\Nss=3$.]{\label{PDPChamb.b}\includegraphics[scale=0.4]{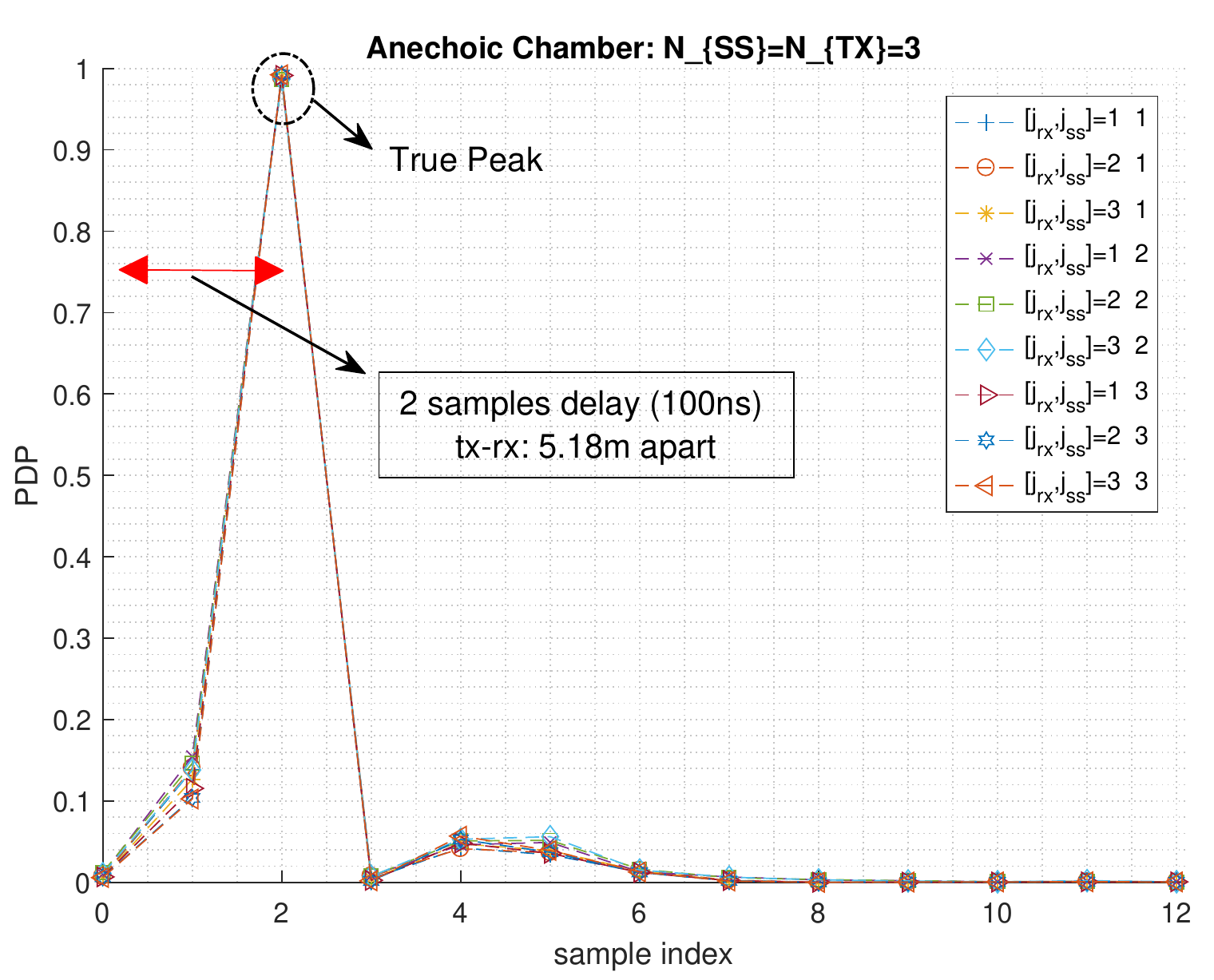}}
\caption{Plots of PDP in anechoic chamber (Fig. \ref{fig12c}). {{For $\Nss=2\rightarrow \SMM\neq \identityMx$ (left), hence, PDP exhibits extra artifact peak. No such peak is observed when $\Nss=3\rightarrow \SMM = \identityMx$. Yet, peaks in both scenarios are shifted by few samples due to the existence of STO, pre-advancement, or CDD.}}}
\label{PDPChamb}
\end{figure}
%
%
%
To better understand the effect of CDD and SMM on range measurement, we performed experiments in an anechoic chamber
(Fig. \ref{fig12c}) wherein $\Nmp\approx 1$ (no multipath). In cases when the CSI matrix is not full rank, i.e. $\Nss\neq \Ntx$, we expect $\SMM^{(\scidx)}\neq \identityMx$.  In this situation, the PDP yields more than one peak $\delDisc_{\rm peak}=\lfloor \Nsc\ScSpacing \DelCoeff_{0}^{\rxidx,\txidx}+{\CDDaCoeff}_{\txidx}+{\CDDbCoeff}_{\rxidx} \rceil$, $\txidx=1\cdots\Ntx$, where ${\CDDaCoeff}_{\txidx}$, ${\CDDbCoeff}_{\rxidx}$ are the cyclic shifts before and after spatial mapping on the $\txidx$th transmit chain and the $\rxidx$ spatial stream. This is indeed the case as shown in Fig. \ref{PDPChamb}. {{Fig. \ref{PDPChamb.a} uses data collected from a setup where transmitter and receiver arrays directly face each other whereas, in Fig. \ref{PDPChamb.b}, the receiver is rotated by 90 degrees. The latter experiment was performed to understand whether we can achieve a full channel matrix ($\Nss =3$) in non-scattering anechoic chamber.}}

 In Fig. \ref{PDPChamb.a}, PDP is plotted for those packets that encounter a channel with $\Nss=2$. As expected, peaks of equal strength is observed (for all transmit-receive sub-channels) which cannot be justified by the echo-free nature of the propagation environment. This is not observed in Fig. \ref{PDPChamb.b} where $\Nss=3$ and the SMM is often non-existent (explained later on). {{Nevertheless, in both figures, peaks are shifted to the right by 2 samples which could be caused by STO, pre-advancement, or CDD.}} \footnote{{{Note that transmitter-receiver are 5.18m apart in anechoic chamber experiment which should produce a peak at sample index "0".}} } 

The conclusion here is that raw CSI is unusable. One has to derive channel-related terms from CSI metrics in order to do positioning, a fact that is often underappreciated  in the field of CSI-based positioning.
%
\subsubsection{Time Domain Windowing}
\begin{figure}[t!]
\centering
\includegraphics[scale=0.8]{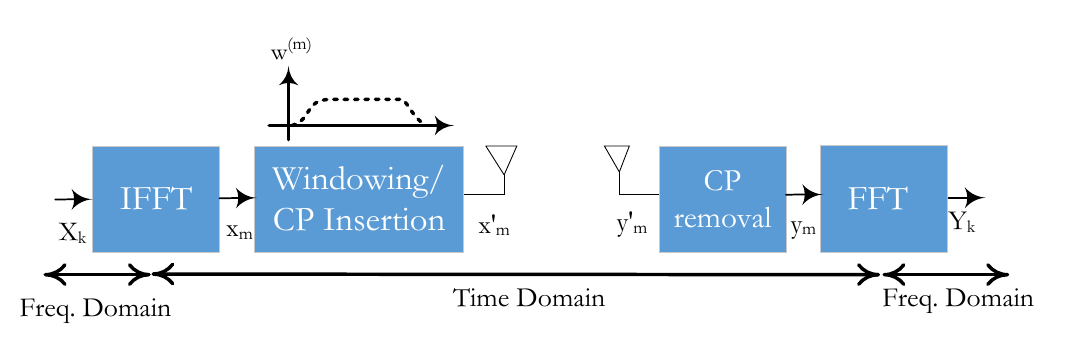}
\caption{{{Illustration of the mixed effect of windowing, CP removal, and guard-band insertion in SISO-OFDM system.}}}
\label{fig.3}
\end{figure}

\begin{figure*}[t!]
\centering
\subfloat[Phase $\angle \mbox{csi}$.]{\label{fig4a}\includegraphics[scale=0.43]{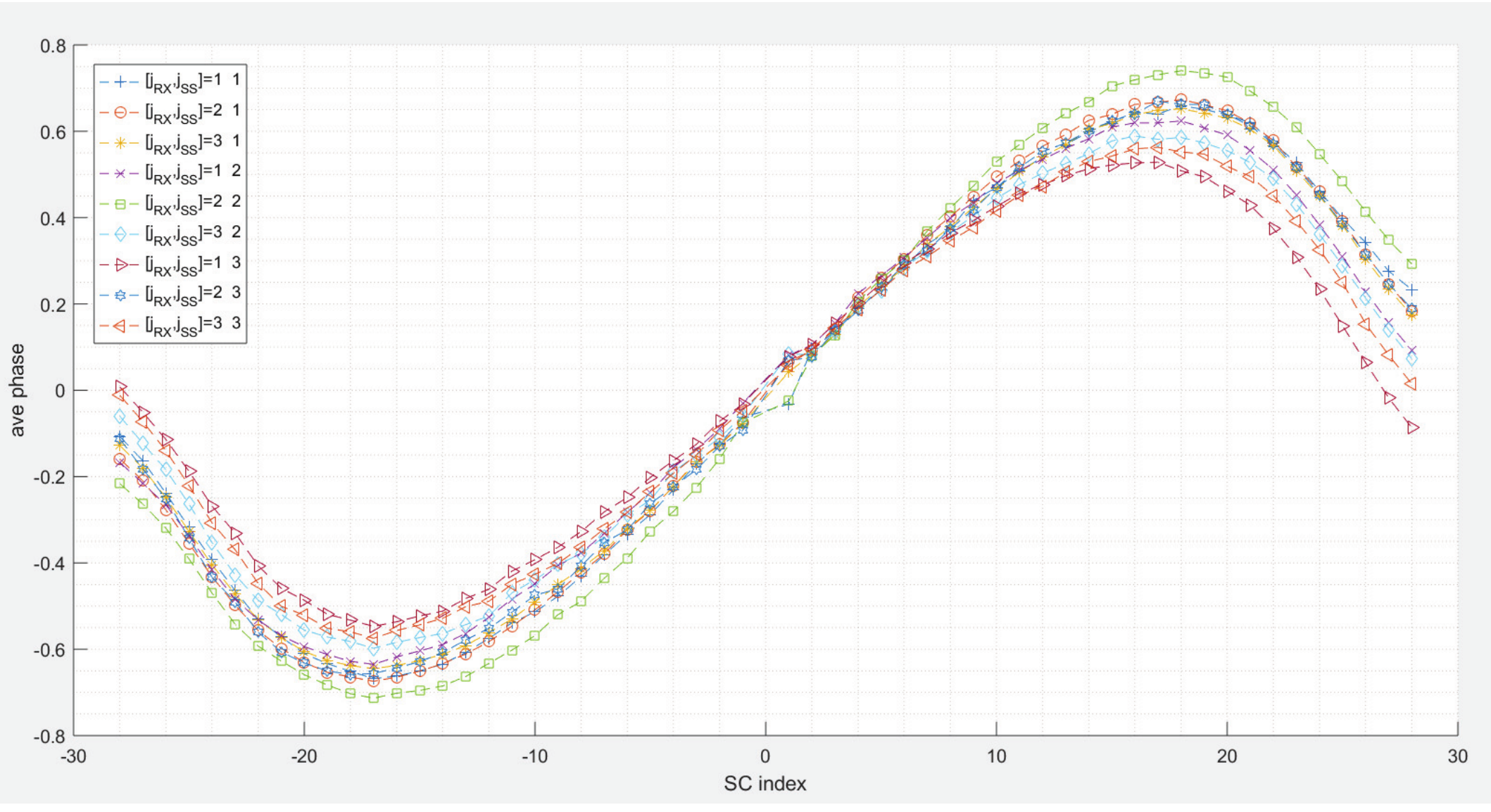}} 
\subfloat[Amplitude $|\mbox{csi}|$.]{\label{fig4b}\includegraphics[scale=0.43]{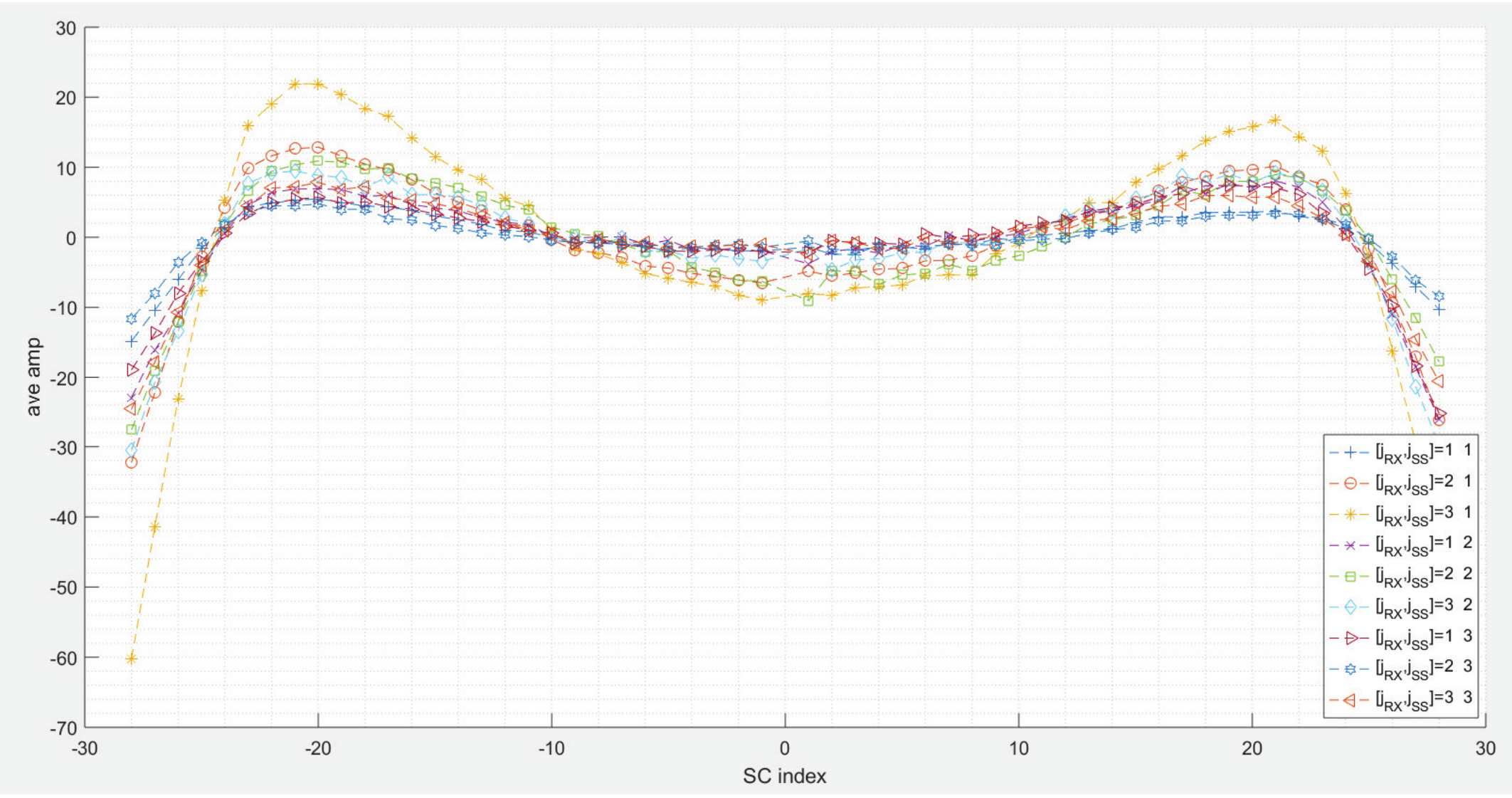}} \\
\subfloat[Comparison of $|\mathcal{F}\{\mbox{csi}\}|$ with truncated Tuckey window (blue curve).]{\label{fig4c}\includegraphics[scale=0.43]{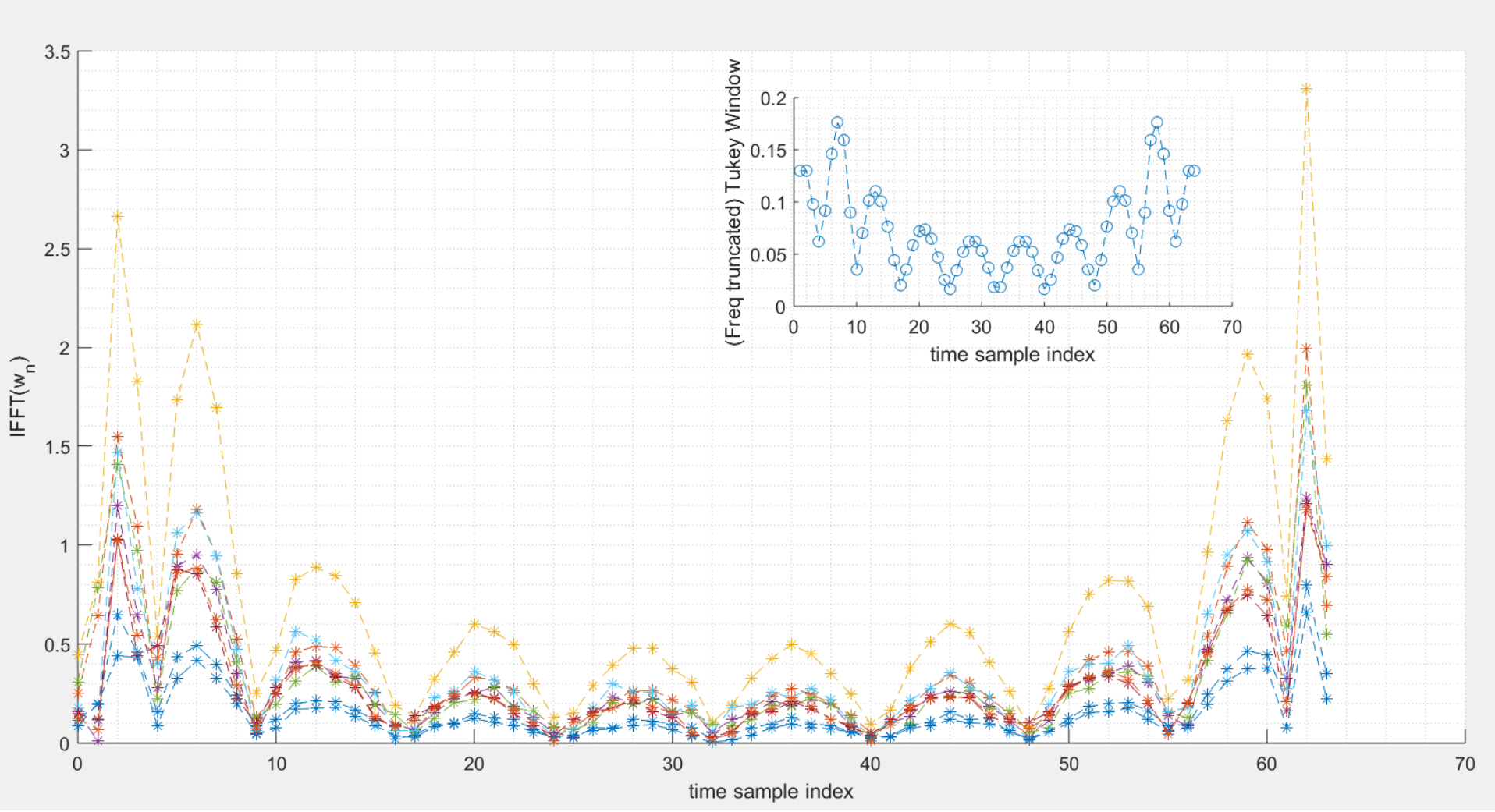}}
\caption{{ Experimental results from conducted test measurements for (a) $\angle \mbox{csi}$, (b) $|\mbox{csi}|$, (c) and $|\mathcal{F}\{\mbox{csi}\}|$ for different $\Nss$ and $\Nrx$. The similarity of the truncated Tucky window (blue curve) to the experimental results in (c) corroborates that the non-linearities observed in (a) and (b) are due to the mixed effect of time-domain windowing, guard-band insertion, and CP removal.}}
\label{fig.4}
\end{figure*}

In examining the CSI obtained in a controlled \textit{conducted} test (Fig. \ref{fig12d}), and in the anechoic chamber, non-linearities of regular shape were observed in both phase and amplitude of CSI as shown in Fig. \ref{fig4a} and Fig. \ref{fig4b}. The symmetric phase and amplitude non-linearity $\WinFreq^{(\scidx)}=\mathcal{F}\{\WinTime^{(\delDisc)}\}$ on CSI (after FFT operation at the receiver) advocates a real-time operation $\WinTime^{(\delDisc)}$ (after IFFT operation at the transmitter). Importantly, this phase distortion can degrade the ranging accuracy. We claim that this effect arises due to the combination of time-domain windowing, cyclic-prefix (CP) removal, and guard-band insertion at the transmitter as shown in Fig. \ref{fig.3} and the logic is as follows: {{Wireless communications systems follow a block-wise design methodology where hierarchies of subsystems\footnote{e.g. scrambling $\rightarrow$ FEC encoding $\rightarrow$ stream parsing $\rightarrow$ interleaving $\rightarrow$ mapper $\rightarrow$ channel $\rightarrow$ equalization $\rightarrow$ de-mapper $\rightarrow$ de-interleaving $\rightarrow$ de-parser $\rightarrow$ FEC decoder $\rightarrow$ de-scrambler} are used at the transmitter and receiver. This approach works because of the linearity of the operation performed in each block, hence, an inner block (say channel-equalization) remains transparent to the outer block (say encoding-decoding). This reversibility is true for most operations along a wireless chain except a few, where CP insertion-removal is the most important one. When CP of the training sequence (from which CSI is calculated) is removed at the receiver, what passes through is a sequence that is windowed (in time domain) from tail but intact from head. That is because the rising head of the time-domain windows are often not long enough to get passed CP and split into the OFDM symbol, but the falling tail of that time-domain window will impact the tail of OFDM symbol. This effect causes the observed distortion.}}

To further investigate this hypothesis, we worked on measurements collected in the \textit{conducted} test setup. In this setting, and based on the model in \eqref{eq.4}, $\Nmp=1$ and $\Nss=\Ntx \rightarrow \SMM=I$, hence CSI with linear phase (vs $\scidx$) was expected, like $\CSIElem_{\rxidx,\ssidx}^{(\scidx)}=\gamma \exp(-{2\pi i\scidx}\zeta/{\Nsc}) +n_{\scidx}$ where $\zeta=\Nsc\ScSpacing\DelCoeff_{0}^{\rxidx,\txidx}+{\CDDaCoeff}_{\txidx}+{\CDDbCoeff}_{\ssidx}$, the latter two terms are the cyclic shifts after and before spatial mapping, $\ScSpacing$ is the OFDM subcarrier spacing, and $\gamma=\beta_0 \exp(-2\pi i\freq_0\DelCoeff_0^{\rxidx,\ssidx})$ is a complex coefficient. Since the non-linearity is completely constant regardless of the choice of attenuators, cable length, etc., it implies a systematic operation happening in hardware. In fact, taking FFT of CSI  yields $\mathcal{F}_{\scidx}^{-1}\{\CSIElem_{\rxidx,\ssidx}^{(\scidx)} \}=\gamma \exp(\zeta_{f})\WinTime^{(\delDisc-\zeta_I)}$ where $\zeta_f$ and $\zeta_I$ are the fractional and integer part of $\zeta$. This time-domain signal is plotted in Fig. \ref{fig4c}. This is a Tukey window as recommended in IEEE 802.11 standard \cite{IEEE80211N}.\footnote{One should note that the Tukey window is a flat function with smooth edge falloff. However, the window we observe through CSI has an FFT whose $\Ngrd/2$ upper (and $\Ngrd/2$ lower) values are zeroed as a result of guard subcarrier exertion, which gives rise to Fig. \ref{fig4c}.} Whereas the results for Atheros 93xx chipset are presented here, the same observation were made for Intel 53xx chipset. In the general case, the CSI model in \eqref{eq.4} is revised as
\begin{equation}\label{eq.5} \small \hspace{-0.2cm}
\begin{aligned}
\CSIElem_{\rxidx}^{(\scidx)}=&\bigg(\sum_{\txidx=1}^{\Ntx}\ChanElem_{\rxidx,\txidx}^{(\scidx)}{\CDDaElem}_{\txidx,\txidx}^{(\scidx)}\sum_{\ssidx=1}^{\Nss}\SMMElem_{\txidx,\ssidx}^{(\scidx)}{\CDDbElem}_{\ssidx,\ssidx}^{(\scidx)} \bigg)\\
& |\tilde{\WinFreq}^{(\scidx)}| e^{i\angle\tilde{\WinFreq}^{(\scidx)}} +n_{\scidx}
\end{aligned}
\end{equation}
where
\begin{equation}\small \nonumber 
\tilde{\WinFreq}^{(\scidx)}=\mathcal{F}\left \{ \WinTime^{(\delDisc)}\cdot \mathrm{rect}_{\Ntot}({\delDisc}/{\Nsc}) \right \}\circledast \mathrm{rect}_{\Nsc}({\scidx}/{\Nnz})
\end{equation}
and $\mathrm{rect}_{\Ntot}({\delDisc}/{\Nsc})$ is a time-domain rectangle function of length $\Ntot=\Nsc+\Ncp$ to represent the CP removal operation on OFDM symbol, $\mathrm{rect}_{\Nsc}({\scidx}/{\Nnz})$ is a frequency-domain rectangle of length $\Nsc=\Nnz+\Ngrd$ to represent guard band insertion operation in OFDM systems, {{and $\WinTime^{(\delDisc)}$ is the time-domain windowing function}}. $\Ncp$, $\Ngrd$, and $\Nsc$ are the length of OFDM cyclic prefix (CP), the number of guard subcarriers, and the total number of subcarriers in OFDM system, respectively. Also $\mathcal{F}(\cdot)$ and $\circledast$ are the FFT and circular convolution operators. Since this is a deterministic effect {{ that stems from a systematic design choice}}, a one-time non-linear fitting to the phase curve in Fig. \ref{fig4a} and de-rotating CSI phase accordingly would be sufficient without any concern with respect to over-fitting.\footnote{Our fit is a 3rd-degree polynomial which resulted in $-7\cdot10^{-5} \scidx^3+3\cdot10^{-5}\scidx^2+0.05\scidx$.}

\textit{\underline{Discussion}}: The existence of phase non-linearity in Fig. \ref{fig.4} has led some researchers to associate this with the I/Q imbalance phenomenon \cite{zhu2018pi}. In several different works, e.g. \cite{yang2013rssi, zhou2014lifi, wu2013csi}, the trigonometric-like shape of the CSI phase (as depicted in Fig. \ref{fig.4}) has led to incorrect representation of CSI as $|\CSIElem^{(\scidx)}| \exp(i\sin(\angle \CSIElem^{(\scidx)}))$. 
The unrecognised, deleterious effects of these baseband operations have led to the belief that CSI is not usable for ToF estimation and made range-based indoor positioning a less fruitful area of investigation. Chronos \cite{vasisht2016} is able to measure ToF by only using the zero subcarriers (at different frequency bands), a workaround that dodges all the deteriorations explained earlier. However, this is not the case if one needs to use CSI on arbitrary set of subcarriers for ToF estimation. On the other hand, estimating AoA using CSI circumvents these problems, as differencing the phases of the CSI at receive antennas eliminates the effect of the aforementioned additive phases imposed at the baseband of the transmitter \cite{kotaru2015, xiong2013, sen2013}. 
\subsection{Impact of Imperfect Signal Processing}

The matrix equation in \eqref{eq.3} assumes perfect synchronization between the transmitter and receiver. Such assumption is not realistic as communication always suffers from lack of perfect time/frequency synchronization. To account for this, the CSI model in \eqref{eq.3} is revised as
\begin{equation}\label{eq.7} \small \hspace{-0.2cm}
\begin{aligned}
\CSIMx^{(\scidx)}(\timeDisc)\leftarrow \CSIMx^{(\scidx)} \underbrace{\begin{bmatrix}
 f(\PsiOne_1(\timeDisc),\PsiOne_2(\timeDisc))&\cdots&0 \\ 
 \vdots & \ddots  &\vdots  \\ 
 0&\cdots   & f(\PsiOne_1(\timeDisc),\PsiOne_2(\timeDisc))
\end{bmatrix}}_{\PsiMx^{(\scidx)}(\timeDisc)}
\end{aligned}
\end{equation}
where $\CSIMx^{(\scidx)}$ is given by \eqref{eq.3},  $\Psi^{(\scidx)}(\timeDisc)$ is an $\Nrx$ by $\Nrx$ matrix of complex and time-dependent elements  $f(\PsiOne_1(\timeDisc),\PsiOne_2(\timeDisc))=\exp(-i(\scidx\PsiOne_1(\timeDisc)+\PsiOne_2(\timeDisc)))$ to account for phenomena such as symbol timing offset (STO), carrier frequency offset (CFO), sampling frequency offset (SFO), and carrier phase offset (CPO). {{Since the chains (transmit and receive) in today’s MIMO systems are driven by one oscillator in an $\Nrx \times \Ntx$ MIMO system, every pair of transmit-receive ports $(n_{\rm rx}, n_{\rm tx})$ observe similar synchronization error in \eqref{eq.7}. Please note the difference between the time index $\timeDisc$ in \eqref{eq.7} (to distinguish CSI for different packets) and delay index $\delDisc$ in \eqref{CIRformula} (to distinguish discrete multipath components of the channel).}}

In general, $\Psi^{(\scidx)}(\timeDisc)$ can be an arbitrary matrix with non-zero elements. However, when there is no coupling between receiver chains, this matrix will be diagonal. Also given that all RF chains in MIMO WLAN systems use a common oscillator/synthesizer, the complex diagonal elements of $\Psi^{(\scidx)}(\timeDisc)$ are the same. Our extensive experiments in the anechoic chamber (Fig. \ref{fig12c}) verifies the following two hypotheses regarding the phase of $f(\cdot,\cdot)$: \textit{(i)} linear in subcarrier index \textit{(ii)} highly variable even in purely static environment. These additive phase terms highly degrade the accuracy of the CSI-based ranging as reported in several localization studies \cite{kotaru2015, xiong2013, vasisht2016} and are discussed next.

\subsubsection{Frequency Errors} In down-converting analog passband (PB) signal to baseband (BB), the following errors are introduced into the CSI:
\begin{itemize}
\item \textbf{CFO/CPO}: The generated carrier at the receiver can be represented by a complex exponential. CFO exists when the receiver's carrier frequency $\freq_0^\prime$ drifts from the transmitted carrier frequency $\freq_0$ by $\CFOoff=\freq_0^\prime-\freq_0$ due to residual errors in receiver's phase locked loop (PLL).\footnote{The CFO can also be due to Doppler effect. Nonetheless, contribution of the latter to $\CFOoff$ is considerably less compared to oscillator frequency mismatch.}

On the other hand, CPO $\CPOoff$ is imposed because receiver's voltage controlled oscillator (VCO) starts from a random phase every time the synthesizer restarts and the phase locked loop (PLL) cannot completely compensate for the phase difference between generated carrier and received signal. Both of these effects are shown to affect CSI in the following manner
\begin{equation}\label{eq.9} \small
\hat{\CSIElem}_{\rxidx,\ssidx}^{(\scidx)}(\timeDisc)=\CSIElem_{\rxidx,\ssidx}^{(\scidx)}(\timeDisc) e^{-2\pi i\left(\dfrac{\CFOCoeff \cdot \CFOCalibFunc(\timeDisc)}{\Nsc} +\CPOoff \right)}+\NoiseSamp_{\scidx}
\end{equation}
where $\CFOCoeff=(\freq_0^{\prime}-\freq_0)/\ScSpacing$ is the CFO normalized with OFDM subcarrier spacing $\ScSpacing$. Equation \eqref{eq.9} signifies an additive phase that is cumulative in time as denoted by $\CFOCalibFunc(\timeDisc)$. Due to its accumulative nature, CFO is regularly tracked by the receiver and compensated for. However, the residual leftover can be detrimental in precise ranging. 
\end{itemize}
\subsubsection{Timing Errors}
\begin{figure}[t!]
\centering
\includegraphics[scale=0.5]{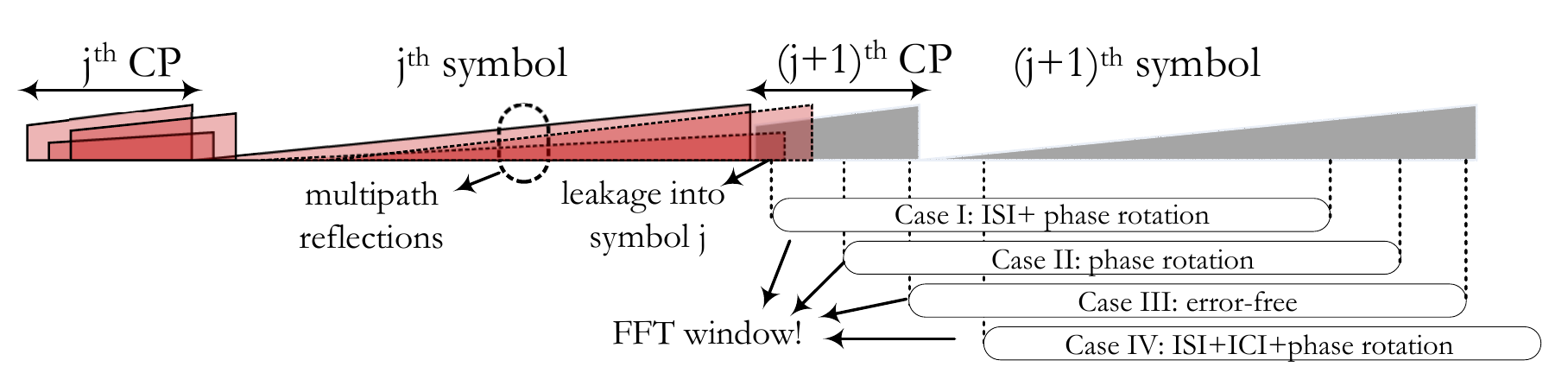}
\caption{Two consecutive OFDM symbols $j$ and $j+1$ are shown with a triangle. Overlapping triangles reflect multipath effect. Different choices of {{OFDM FFT window (a.k.a. symbol boundary)}} lead to four types of deteriorations in detecting symbol $j+1$.}
\label{fig.6}
\end{figure}

These errors happen when receiver (transmitter) samples (synthesizes) signals at mismatching rates. There is also the significant issue of symbol boundary detection as discussed next:

\begin{itemize}
\item \textbf{SFO}: In modern homodyne architectures, the same oscillator triggering the mixer drives the analog-to-digital converter (ADC). If the ADC samples the received signal with rate $\SampInt^{\prime}$ different from transmitter's synthesization rate
 $\SampInt$, SFO is experienced. This is manifested as an additive phase shift proportional to the subcarrier index and cumulative in time \cite{schmidl1997robust, speth1999optimum}. Mathematically,
\begin{equation}\label{eq.10}
\hat{\CSIElem}_{\rxidx,\ssidx}^{(\scidx)}(\timeDisc)=\CSIElem_{\rxidx,\ssidx}^{(\scidx)}(\timeDisc) e^{-2\pi i\scidx\left(\dfrac{\SFOCoeff \cdot \SFOCalibFunc(\timeDisc)}{\Nsc}\right)}+\NoiseSamp_{\scidx}
\end{equation}
where $\SFOCoeff=(\SampInt'-\SampInt)/\SampInt$ is the SFO normalized with the sampling time and  $\SFOCalibFunc(\timeDisc)$ denotes the SFO calibration interval.

\item \textbf{STO}: STO is the most degrading effect arising due to the lack of knowledge about the beginning of the received OFDM symbol \cite{speth1999optimum}. This uncertainty emerges as it is not a-priori known when to expect a packet. Since OFDM systems function on blocks of (time domain) samples, named symbols, it is crucial that the right block is fed to the FFT demodulator. To find out about the symbol boundary, header starts with known, periodic sequences (named short-training fields-STF) and auto-correlator/cross-correlator is utilized at the receiver to capture and detect the presence of WiFi signals. However, because of the length limitations of these sequences, error in determining symbol boundary cannot be fully eliminated leading to irreversible errors such as inter-carrier interference (ICI), inter-symbol interference (ISI), and phase rotation, as seen in Fig. \ref{fig.6}.\footnote{{{ISI is experienced in case I of Fig. \ref{fig.6} because there is multipath leakage from $j$th symbol into the FFT window of the $j+1$th symbol. This is different from Case IV where not only leakage from the next symbol (i.e. j+2 which is not plotted) causes ISI, but there is ICI as well since the FFT window is missing the beginning of OFDM frame. To summarize, FFT window should neither advance too much into CP (to avoid ISI with the previous symbol) nor should it progress into main part of OFDM symbol (to avoid ICI and ISI with the next symbol).}} } This  phase rotation can be shown to impact CSI in the following manner:
\begin{equation}\label{eq.sto}
\hat{\CSIElem}_{\rxidx,\ssidx}^{(\scidx)}(\timeDisc)=\CSIElem_{\rxidx,\ssidx}^{(\scidx)} e^{-2\pi i\left(\dfrac{\scidx\STOcount(\timeDisc)}{\Nsc}\right)}+\NoiseSamp_{\scidx}
\end{equation}
\item \textbf{OFDM Pre-advancement}: Accounting for STO uncertainty, and to avoid irrevocable ICI/ISI, almost all NIC chipsets intentionally (upon estimating symbol boundary) borrow $\Preadvancement$ samples from current OFDM symbol's CP. This operation, named pre-advancement, guarantees that FFR input samples are ISI/ICI free, and only (clockwise) cyclically shifted (Case II in Fig. \ref{fig.6}) which creates phase rotation after FFT given by:\footnote{pre-advancement won't impact decoding quality as both payload and channel estimation (HT-LTF) symbols undergo the same shift, hence equalization removes it.} 
\begin{equation}\label{eq.11}
\hat{\CSIElem}_{\rxidx,\ssidx}^{(\scidx)}=\CSIElem_{\rxidx,\ssidx}^{(\scidx)} e^{-2\pi i\left(\dfrac{\scidx\Preadvancement}{\Nsc}\right)}+\NoiseSamp_{\scidx}
\end{equation}
\textit{\underline{Discussion}}: Positioning based on the unprocessed CSI will be severely impacted as $\STOcount+\Preadvancement=1$ will cause $15$m ranging inaccuracy at best. This is evident from our experimental measurements in Fig. \ref{PDPChamb}: Whereas in the chamber the transmitter and receiver were $5$m  apart, calling for a PDPs that climax at the very first sample ($\timeDisc=0$), the true peak actually happens at the third sample, an anomalous behaviour that is a testimony to the deliberate clockwise (left) cyclic shifting of OFDM symbol.
\end{itemize}

Accounting for non-idealities due to AGC, CFO, CPO, SFO, STO, and pre-advancement, the CSI model is revised as follows
{{
\begin{equation}\label{eq.12} \small
\begin{aligned}
\hat{\CSIElem}_{\rxidx,\ssidx}^{(\scidx)}(\timeDisc)=
&
\alpha_{\rm agc}\CSIElem_{\rxidx,\ssidx}^{(\scidx)}\underbrace{\left(e^{-2\pi i\CPOoff}\right)}_{\rm CPO}\underbrace{\left(e^{-\frac{2\pi i\CFOCoeff \CFOCalibFunc(\timeDisc)}{\Nsc}}\right)}_{\rm CFO} \\ &
\times\underbrace{\left(e^{-\frac{2\pi i\scidx\SFOCoeff \SFOCalibFunc(\timeDisc)}{\Nsc}}\right)}_{\rm SFO}\underbrace{\left(e^{-\frac{2\pi i\scidx(\STOcount+\Preadvancement)}{\Nsc}}\right)}_{\rm STO + pre-advancement}+\mathcal{J}_{\scidx}
\end{aligned}
\end{equation}
where, according to Eq. \eqref{eq.4}, $\CSIElem_{\rxidx,\ssidx}^{(\scidx)}$ is given by
\begin{equation} \nonumber
\CSIElem_{\rxidx,\ssidx}^{(\scidx)}= \sum_{\txidx=1}^{\Ntx}\ChanElem_{\rxidx,\txidx}^{(\scidx)}{\CDDaElem}_{\txidx,\txidx}^{(\scidx)}\SMMElem_{\txidx,\ssidx}^{(\scidx)}{\CDDbElem}_{\ssidx,\ssidx}^{(\scidx)}+\NoiseSamp_{\scidx}
\end{equation}
}}
The additive term $\mathcal{J}_{\scidx}$ entails noise $\NoiseSamp_{\scidx}$, ISI, and ICI. Despite its sophisticated look, the multiplicative error terms in \eqref{eq.12} can be compactly represented by $\exp(-i(\scidx\PsiOne_1(\timeDisc)+\PsiOne_2(\timeDisc)))$ as initially claimed in \eqref{eq.7}.

\section{CSI Calibration} \label{CSIcalib}
We have discussed so far that ranging based solely on CSI is a futile effort unless \textit{(i)} the effect of deterministic SMM, CDD, and mixed windowing operations are cancelled out \textit{(ii)}  random phase errors due to the lack of synchronization are compensated for. 

In the following, we investigate the statistical behaviour of the CSI random phase errors and introduce techniques to remove them. Our goal is to estimate synchronization errors in \eqref{eq.12} in the aforementioned onerous problem where errors are changing from packet to packets, thus, rendering classic estimation (ML, MMSE, etc.) approaches that rely on availability of many samples unusable.

\subsection{Statistical Error Characterization} \label{CSIcalib:StatisticalAnalysis}
Due to the highly volatile nature of phase errors, differencing across time keeps the volatility while eliminating stagnant channel terms.\footnote{As a rough figure, the parameters of the indoor wireless channel change in the order of tens of ms.} Doing so for consecutive CSI samples and performing phase unwrapping (w.r.t the subcarrier index $\scidx$) yields\footnote{One has to be wary of the fact that we do not get to observe $\Delta\PsiOne_1\scidx+\Delta\PsiOne_2$ but its $2\pi$  modulus.}
\begin{equation}\label{eq.14} \small
\begin{aligned}
&\underbrace{{\rm uwrp}_{\scidx}\left[ \angle \CSIElem_{\rxidx,\ssidx}^{(\scidx)}(\timeDisc_1)-\angle \CSIElem_{\rxidx,\ssidx}^{(\scidx)}(\timeDisc_2)\right]}_{\Delta(\angle \CSIElem^{(\scidx)})(\timeDisc_1,\timeDisc_2)}
\\&
=
{\rm uwrp}_{\scidx}[\underbrace{(\PsiOne_1(\timeDisc_1)-\PsiOne_1(\timeDisc_2))}_{\Delta\PsiOne_1(\timeDisc_1,\timeDisc_2)}\scidx+\underbrace{(\PsiOne_2(\timeDisc_1)-\PsiOne_2(\timeDisc_2))}_{\Delta\PsiOne_2(\timeDisc_1,\timeDisc_2)} ~ ({\rm mod ~} 2\pi)]
\\&
=
{\rm uwrp}_{\scidx}[\Delta\PsiOne_1(\timeDisc_1,\timeDisc_2)\scidx ~({\rm mod ~} 2\pi)]+\Delta\PsiOne_2(\timeDisc_1,\timeDisc_2) ~ ({\rm mod ~} 2\pi)
\end{aligned}
\end{equation}
where ${\rm uwrp}_{\scidx}[\cdot]$ is phase unwrapping w.r.t to $\scidx$ {{and $\timeDisc_1$ and $\timeDisc_2$ are arbitrary time indices with the constraint that $(|\timeDisc_2-\timeDisc_1| \SampInt < T_{\rm c})$ with $T_{\rm c}$ being the coherence time of the channel.}} Also, $x ~({\rm mod ~} 2\pi)$  is the modulo operation, which is denoted by $[x]_{2\pi}$,  hereinafter. To gain insights into the statistical nature of $\PsiOne_{1(2)}(\timeDisc)$, we use the measurements collected in an anechoic chamber.  Fig. \ref{fig8a} shows CSI phase difference vs. subcarrier index for two cases: \textit{(i)}  $\Npkt=12$  \textit{(ii)}  $\Npkt=8000$ CSI measurements. Fig. \ref{fig8b} displays the empirical PDF of $\Delta\PsiOne_{\freeidx}$, $\freeidx=\{1,2\}$, for $\Npkt=8000$. The following conclusions are drawn:
\begin{figure*}[t!]
\centering
\subfloat[CSI phase (simulation).]{\label{fig8aa}\includegraphics[scale=0.35]{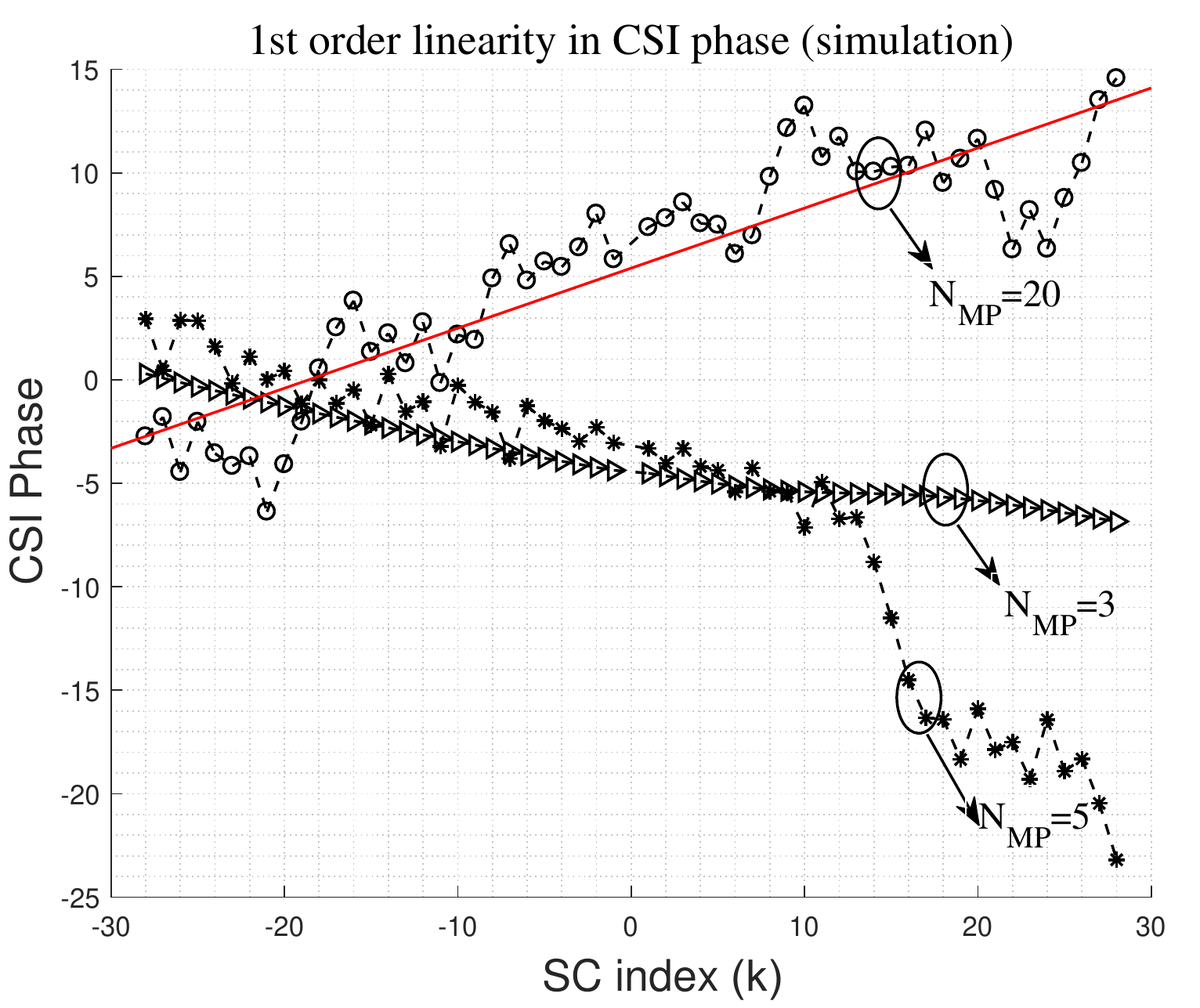}} 
\subfloat[CSI phase difference for 8000 packets (experiment).]{\label{fig8a}\includegraphics[scale=0.4]{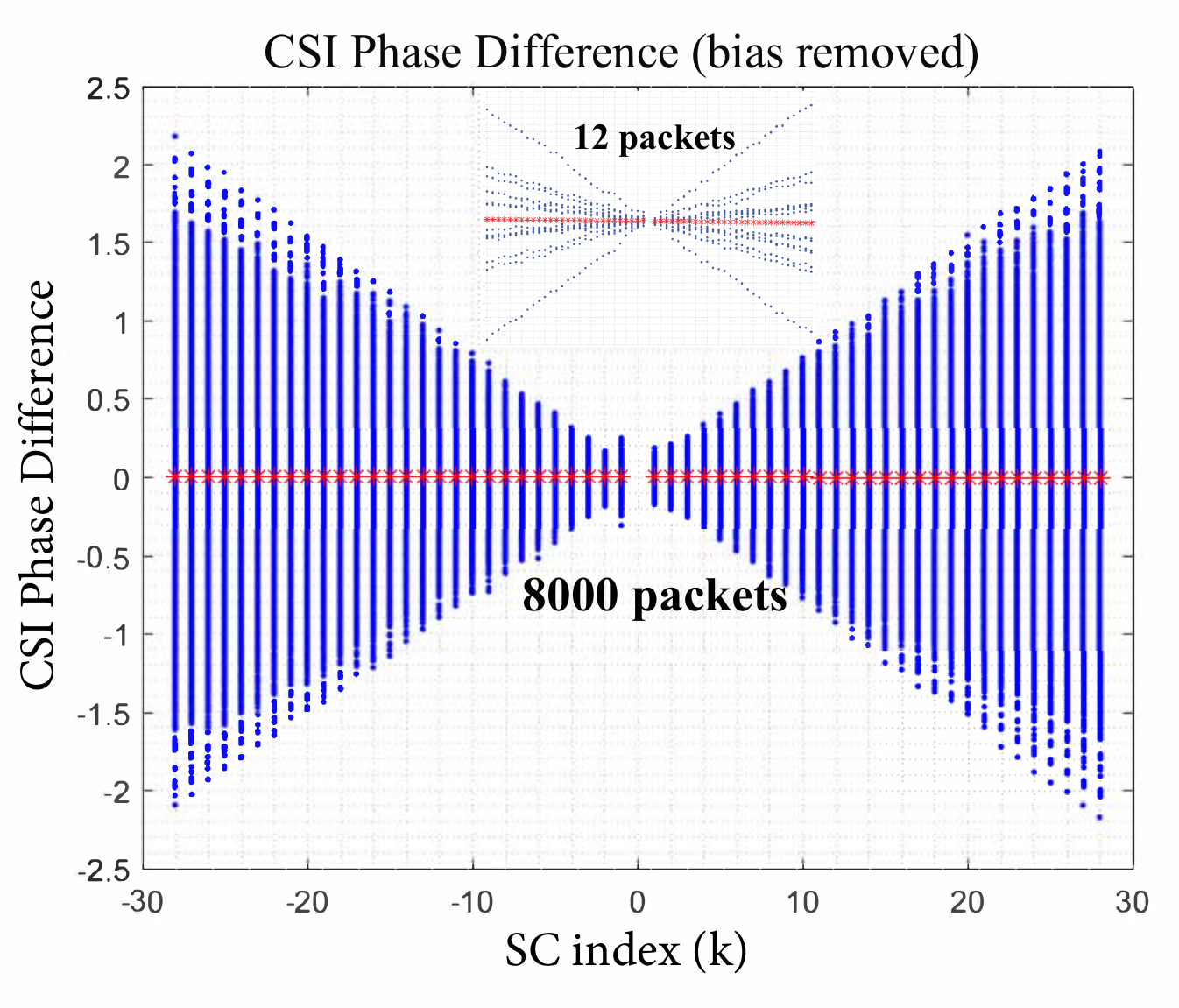}}
\subfloat[Histogram of $\Delta\PsiOne_{\freeidx}$, $\freeidx=\{1,2\}$ for 8000 packets (experiment).]{\label{fig8b}\includegraphics[scale=0.35]{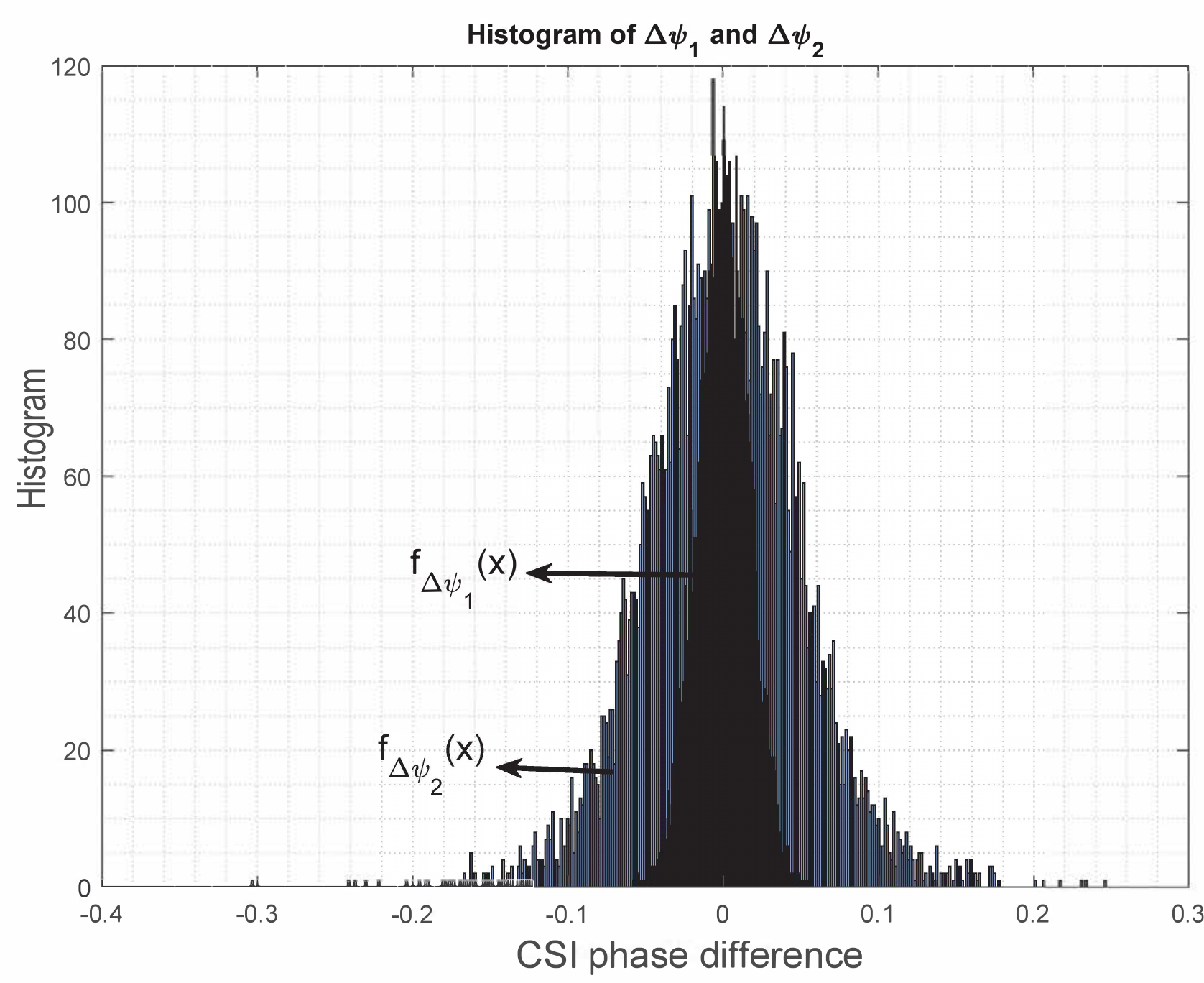}} 
\caption{{{(a) simulation: This figure proves that channel phase has affine component (red straight line) $\Nmp = 3, 5, 20$. (b) experiment: differencing phase of CSI of consecutive packets in \eqref{eq.12} eliminates channel component while keeping the volatile synchronization component. This figure also proves that  linearity (vs. $\scidx$) is a valid assumption for $\Delta(\angle \CSIElem^{(\scidx)})$. Moreover, $\Delta(\angle \CSIElem^{(\scidx)})$ is a zero mean random variable whose mean is shown by the red horizontal line. (c) experiment: histogram of $\Delta\PsiOne_{\freeidx}$, $\freeidx=\{1,2\}$ corroborates the validity of Gaussian assumption.}}}
\label{fig.8}
\end{figure*}
\begin{itemize}
\item Even for as low as $\Npkt=12$, the randomness introduced by $\PsiOne_{\freeidx}(\timeDisc), r=\{1,2\}$ is so large that it drives the average phase difference (horizontal red line) to zero. This observation substantiates that both $[\Delta\PsiOne_{1}(\timeDisc_1,\timeDisc_2)]_{2\pi}$ and $[\Delta\PsiOne_{2}(\timeDisc_1,\timeDisc_2)]_{2\pi}$ are zero mean random processes.
\item The obvious linearity in Fig. \ref{fig8a} conforms with the derivations in \eqref{eq.12} as was reported in earlier works \cite{xie2015precise}.
\item The drastic changes of $\Delta(\angle \CSIElem^{(\scidx)})=[k\Delta \PsiOne_1+\Delta \PsiOne_2]_{2\pi}$ is because of two effects: (a) The high dynamicity of receiver's synchronization algorithms (b) the $\angle (\cdot)$ operation which delivers not the true angle but the wrapped-around version of it. 
\item  The Gaussianity of  $\Delta\PsiOne_{\freeidx}$, $\freeidx=\{1,2\}$ is proved as follows: Since $\Delta(\angle \CSIElem^{(\scidx)})$ is a Gaussian process (per our observation),  $\Delta(\angle \CSIElem^{(\scidx=0)})=\Delta\PsiOne_{2}$ is Gaussian random variable. Noting that $\Delta\PsiOne_{1}\perp \Delta\PsiOne_{2}$, then ${\varphi}_{\scidx_0 \Delta\PsiOne_{1}}(t)\cdot{\varphi}_{\Delta\PsiOne_{2}}(t)={\varphi}_{\Delta(\angle \CSIElem^{(\scidx_0)})}(t)$, where $\varphi_{a}(t)$ is the characteristic function of random variable $a$. Subsequently, the PDF of $\Delta\PsiOne_{1}$ is attained using the Fourier transform, that is, $f_{\Delta\PsiOne_{1}}=1/\scidx_0 \mathcal{F}\{{\varphi}_{\Delta(\angle \CSIElem^{(\scidx_0)})}(t)/{\varphi}_{\Delta\PsiOne_{2}}(t)\}$ which can be shown to be a Gaussian. This is shown in Fig. \ref{fig8b}.
\item Finally, the knowledge of $\Delta\PsiOne_{\freeidx}=\mathcal{N}(0,\sigma_{\freeidx}^2)$ implies  $\PsiOne_{\freeidx}=\mathcal{N}(\mu_{1(2)},\sigma_{\freeidx}^2/2)$, $\freeidx=\{1,2\}$. This is true since the process $\PsiOne_{\freeidx}(\timeDisc)$ has the same distribution for different $\timeDisc$. Yet, so long as the cyclic-prefix (CP) pre-advancement is performed at the receiver, deeming $\PsiOne_{\freeidx}(\timeDisc)$ as a zero-mean random variable \cite{xie2015precise} yields completely biased range estimates. 
\end{itemize}

\textit{\underline{Discussion}}: These findings contradict some views on the uniformity of distributions  $\PsiOne_{\freeidx}$ \cite{Ansari2009,Weeraddana2009}, an assertion seemingly made due to equating the statistical behaviour of $[\PsiOne_{\freeidx}]_{2\pi}$ with that of $\PsiOne_{\freeidx}$.

\subsection{Estimating STO and SFO}\label{CSIcalib:STOSFOestimation}
The unpredictability of phase errors $\PsiOne_{1}(\timeDisc)$ in \eqref{eq.12} stems from the following reasons:
\begin{itemize}
\item Randomness in $\SFOCalibFunc(\timeDisc)$ due to the opportunistic nature of WLAN access protocol.
\item Randomness in $\SFOCalibFunc(\timeDisc)$ due to receiver's ability to initiate calibration using any packet header on the air regardless of whether it was destined to it or not.
\item Errors in estimating the amount of drift $\SFOCoeff$ which depends on how badly the calibrating header is influenced by small scale fading.
\item Errors in estimating the symbol boundary and $\STOcount$ which, again, depends on the fading nature of the channel.
\item OFDM pre-advancement \cite{perahia2013next}.
\end{itemize}
For these reasons, $\PsiOne_{1}(\timeDisc)$ is decorrelated for different $\timeDisc$.  Therefore, only CSI across frequency and space can be used to estimate $\PsiOne_{1}(\timeDisc)$. With this knowledge and given the linearity of the additive phase (in $\scidx$) in \eqref{eq.12}, several previous works \cite{kotaru2015, Sen2010, Xiong2015ToneTrack} adopted a simple CSI phase de-trending to eliminate $\PsiOne_{1}(\timeDisc)$. This estimator can more generally be expressed as  
\begin{equation} \label{eq.est1} \small \hspace{-0.2cm}
\begin{aligned}
\hat{\PsiOne}_1(\timeDisc)= -\angle \frac{1}{(\Nnz-1)} \sum_{\scidx=-\Nnz/2+1}^{\Nnz/2}\left(\CSIElem^{(\scidx)}_{\rxidx,\ssidx}(\timeDisc) {\CSIElem^{(\scidx-1)}_{\rxidx,\ssidx}}^{*}(\timeDisc)\right)
\end{aligned}
\end{equation}
where $\CSIElem_{\rxidx,\ssidx}^{(\scidx)}$ is the $(\rxidx,\ssidx)$th element of the CSI matrix, $\Nnz=\Nsc-\Ngrd$ is the number of non-zero subcarriers and $\Ngrd$ is the number of guard subcarriers at both ends of spectrum that are not used to modulate any symbol. This is an exact estimator, i.e. $\PsiOne_1(\timeDisc)=\hat{\PsiOne}_1(\timeDisc)$, only when \textit{(i)} the channel does not change variably between two adjacent subcarriers, that is $\ChanElem^{(\scidx)}_{\rxidx,\txidx}{\ChanElem^{(\scidx-1)}_{\rxidx,\txidx}}^{*}\approx |\ChanElem^{(\scidx)}_{\rxidx,\ssidx} |^2$ and \textit{(ii)} $\mathbb{E} \{\PsiOne_1(\timeDisc)\}= 0$. 

\textit{None of these two conditions is satisfied in reality}:  As shown in Fig. \ref{fig8aa}, the true channel phase normally has a \textit{first-order linearity}, hence, \eqref{eq.est1} estimates $\PsiOne_{1}$ plus the linear phase term in $\ChanElem_{\rxidx,\txidx}$, which is denoted by $\ChanPhLin_{\rxidx,\txidx}$ hereinafter. In this situation, \eqref{eq.est1} becomes (often negatively) a biased estimator and compensating CSI using it (as in \eqref{eq.20}) gravely impacts ranging accuracy possibly worse than keeping STO and ranging with the original CSI.
The performance of \eqref{eq.est1} is studied for thousands of channel realizations and for two different STO+SFO drift. The bias of the estimator, caused by eliminating the first-order channel linearity $\scidx\cdot \ChanPhLin_{\rxidx,\txidx}$ was observed.
\subsubsection{Alternative Estimators}
In obtaining $\hat{\PsiOne}_1(\timeDisc)$, the following estimator was proven more effective in reducing the estimation error in lieu of \eqref{eq.est1}. 
\paragraph{Spatial/spectral Averaging} Given that all transmit/receive sub-channels experience the same hardware error, averaging can be performed in those dimensions as follows:
\begin{equation} \label{eq.est2} \small
\begin{aligned}
\hat{\PsiOne}_1^{\rm (I)}(\timeDisc)=
&
-\angle\bigg(\frac{1}{(\Nnz-1)\Nrx\Nss}\cdot \\&
\sum_{\scidx=-\Nnz/2+1}^{\Nnz/2}\sum_{\rxidx=1}^{\Nrx}\sum_{\ssidx=1}^{\Nss}\CSIElem_{\rxidx,\ssidx}^{(\scidx)}(\timeDisc) {\CSIElem_{\rxidx,\ssidx}^{(\scidx-1)}}^{*}(\timeDisc)\bigg)    
\end{aligned}    
\end{equation}

Having obtained $\hat{\PsiOne}_1$, compensation is performed with simple post multiplication with the CSI matrix as follows,
\begin{equation}\label{eq.20}
\tilde{\CSIMx}^{(\scidx)}(\timeDisc)=e^{i\left(\hat{\PsiOne}_{1}(\timeDisc)\right)\scidx}\cdot \CSIMx^{(\scidx)}(\timeDisc)
\end{equation}
\begin{algorithm} \scriptsize
\caption{1st-Order Linearity Removal}\label{FirstOrderLinearity}
    \hspace*{\algorithmicindent} \textbf{Input:} $\CSIMx^{(1:\Nsc)}(1:\Npkt)$
    \\
    \hspace*{\algorithmicindent} \textbf{Output:} $\CSIMx^{(1:\Nsc)}(1:\Npkt)$
\begin{algorithmic}
	\Procedure{CompensationI}{\textbf{Input}} 
	\State {{De-rotate $\angle \CSIMx^{(1:\Nsc)}(1:\Npkt)$ by $-7e-5\scidx^3+3e-5\scidx^2+0.05\scidx$ \Comment{Windowing/CP removal effect compensation}}}
	\State $s=0$, ${\rm flag}=0$;
	\State $\mathcal{A}:$ \Comment{Return here to compensate for channel-linearity}
	\For {$\pktidx=1 :  \Npkt$}
    	\If{${\rm flag}==0$}	
	
			\State Estimate $\hat{\PsiOne}_1(\pktidx)$  {using \eqref{eq.est1} or \eqref{eq.est2}};
			\State $s=s+\hat{\PsiOne}_1(\pktidx)$;
		\Else
	    	\State  $\tilde{\PsiOne}_{1}(\pktidx) \approx \hat{\PsiOne}_{1}(\pktidx)-\ChanPhLin_{\rxidx,\txidx}$;
	    	\State $\CSIMx^{(1:\Nsc)}(\pktidx) ={\rm Derotate}(\CSIMx^{(1:\Nsc)}(\pktidx),\tilde{\PsiOne}_{1}(\pktidx))$;			\\
				\Comment{De-rotate CSI phase using \eqref{eq.20}}
		\EndIf
	\EndFor
	\State $\ChanPhLin_{\rxidx,\txidx}=s/\Npkt$; \\
	\State ${\rm flag}=1$; 
 	\State ${\bf Goto}$  $\mathcal{A}$ \Comment{Return to $\mathcal{A}$ only once}
 	\EndProcedure
\end{algorithmic}
\end{algorithm}
\subsubsection{Recovering Channel Phase Linearity}
The goal is to subtract the first-order channel linearity $\ChanPhLin_{\rxidx,\txidx}$ that is removed (along with phase errors) in \eqref{eq.est1}-\eqref{eq.est2}. However, unless the true attenuations $\boldsymbol {\AttCoeff}$, path delays $\boldsymbol{\DelCoeff}$, and $\Nmp$ are precisely known in advance (which is actually the ultimate goal of positioning), no deterministic approach can find $\ChanPhLin_{\rxidx,\txidx}$. Yet, with the knowledge that the channel-related term remains constant over the course of several packets, and leveraging the randomness in $\PsiOne_1$, Algorithm \ref{FirstOrderLinearity} is used to remove the volatility contributed by SFO and find $\tilde{\PsiOne}_{1}(\timeDisc)$. This is corroborated when observing that $\tilde{\PsiOne}_{1}(\timeDisc)$ is closely independent of  $\rxidx$ and $\txidx$ (which is expected as per \eqref{eq.12}) whereas $\hat{\PsiOne}_{1}(\timeDisc)$ varies across antennas.  

\subsubsection{STO Removal}
The previous procedure designed to recover the channel linearity is incapable of eliminating the STO phase. This is because $\STOcount$ varies in much longer time-scale, hence, is somewhat fused into the channel phase $\ChanElem_{\rxidx,\txidx}$. STO manifests itself as jumps at the end of PDP due to cyclic-shifting of CIR. {{This is because any phase shift due to STO in frequency domain ($\scidx$) causes circular rotation by the same amount in time domain ($\timeDisc$). Since, in indoor environments, transmitter-receiver are only several meters away and that bandwidth is limited (20MHz in IEEE802.11n), the first expected peak of true channel CIR (due to LoS arrival) often happens at  $\timeDisc$=0 or 1 which means that any $\STOcount>1$ causes that peak to appear at the end of the PDP (due to circular shift property). }}

 This observation forms the basis to estimate $\STOcount$ through the following logic: The discrete CIR in \eqref{CIRformula} is a linear combination of shifted discrete Dirichlet functions. This is a periodic function with fundamental period $\Nsc$ that varies smoothly from one sample to the next. Therefore, jumps that are observed at the far-end of the PDP due to STO, can be detected and compensated for using Algorithm \ref{STO}.
\begin{algorithm} \scriptsize
\caption{STO Removal}\label{STO}
    \hspace*{\algorithmicindent} \textbf{Input:} $\CSIMx^{(1:\Nsc)}(1:\Npkt)$ outputted from Alg. \ref{FirstOrderLinearity}
    \\
    \hspace*{\algorithmicindent} \textbf{Output:}  $\CSIMx^{(1:\Nsc)}(1:\Npkt)$
\begin{algorithmic}
\Procedure{CompensationII}{\textbf{Input}} 
\State $y=[]$, ${\rm flag}=0$;
\State $\mathcal{A}:$ \Comment{Return here to compensate for STO} 
\For {$\pktidx=1 : \Npkt$} 
	\For {$\rxidx=1 : \Nrx$} 
		\For {$\ssidx=1 : \Nss$} 
			\If {${\rm flag}==0$}
				\State $\CirVec_{\rxidx,\ssidx}^{(\timeDisc)}(\pktidx)=\mathfrak{F}_k^{-1}\{\CSIVec_{\rxidx,\ssidx}^{(\scidx)}(\pktidx) \}$; \Comment{Take IFFT ($\mathfrak{F}_k^{-1}$) of CSI}
				\State ${\pdp}_{\rxidx,\ssidx}^{(\timeDisc)}(\pktidx)=|\CirVec_{\rxidx,\ssidx}^{(\timeDisc)}(\pktidx)|^{2}$;
			
				\State ${\bf d}^{(\timeDisc)}={\rm PeakFinder}(\pdp_{\rxidx,\ssidx}^{(\timeDisc)}(\pktidx))$; \Comment{Find peaks of PDP}
				\State $s=s\cup{\bf d}^{(\timeDisc)}$;
			\Else
				\State ${\rm Derotate}(\CSIVec_{\rxidx,\ssidx,\pktidx}^{(\scidx)},-2\pi\STOcount/\Nsc)$;			\\
				\Comment{De-rotate CSI phase using \eqref{eq.20}}
			\EndIf
		\EndFor
	\EndFor
\EndFor
\State $\STOcount={\rm Mode}(s)$; \Comment{Find the most frequent cyclic shift}
\State ${\rm flag}=1$;
\State ${\bf Goto}$  $\mathcal{A}$ \Comment{Return to $\mathcal{A}$ only once}
\EndProcedure
\end{algorithmic}
\end{algorithm}
\subsection{Removing CFO and CPO} \label{CSIcalib:CFOCPOremoval}
Removing the linear phase terms produced by STO/SFO leaves CFO/CPO errors in \eqref{eq.12} intact. As explained earlier, and similar to SFO, CFO is an accumulative error that has to be tracked by the receiver and compensated for. However, this compensation is crude and leaves behind some residual phase $\PsiOne_2(\timeDisc)$ on CSI. Estimating the latter is not an easy task. That is because:
\begin{itemize}
\item Similar to $\SFOCalibFunc(\timeDisc)$, not much is known about the calibration intervals $\CFOCalibFunc(\timeDisc)$ within the receiver.
\item Small-scale fading highly deteriorates CFO compensation performed at the receiver using high-throughput short-training field (HT-STF).
\item There is no differentiating dimension (as was $\scidx$ in previous case) to distinguish the latter from the channel term.
\item Whereas it was shown earlier that $\PsiOne_2(\timeDisc)$ has Gaussian distribution, it is not this variable that we observe but its wraparound version $[\PsiOne_2(\timeDisc)]_{2\pi}=2\pi\CPOoff+[{2\pi}\CFOCalibFunc(\timeDisc)\CFOCoeff/{\Nsc}]_{2\pi}$.
\end{itemize}

\begin{figure*}[t!]
\centering
\subfloat[LoS (corridor).]{\label{fig12a}\includegraphics[width=4cm,keepaspectratio]{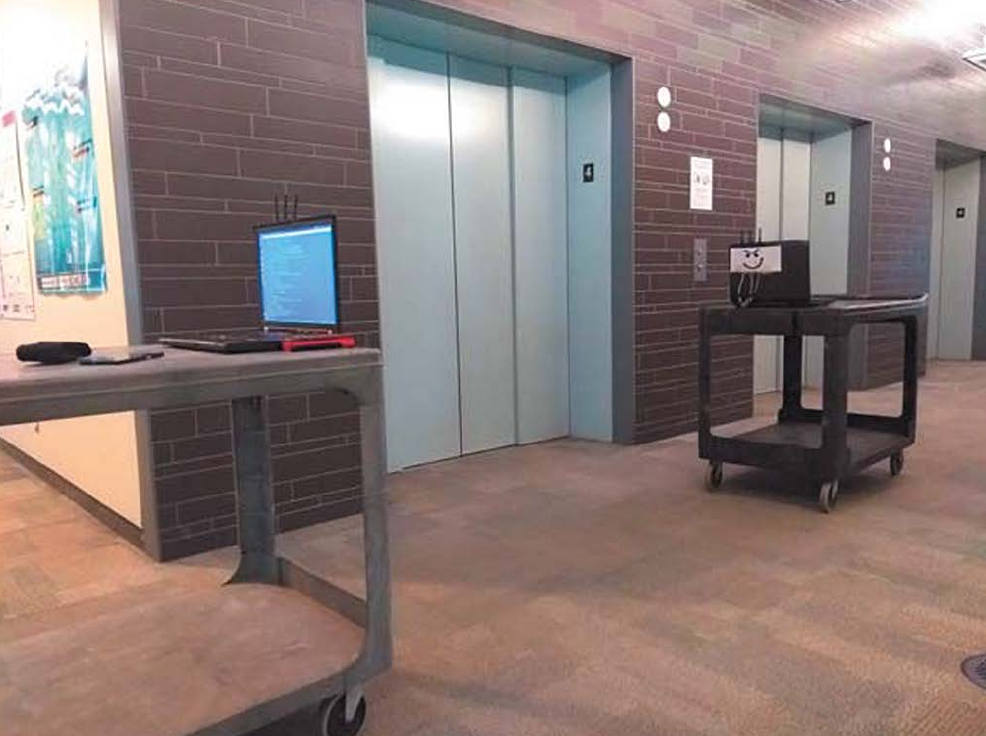}}
\subfloat[NLoS (corridor).]{\label{fig12b}\includegraphics[width=4cm,keepaspectratio]{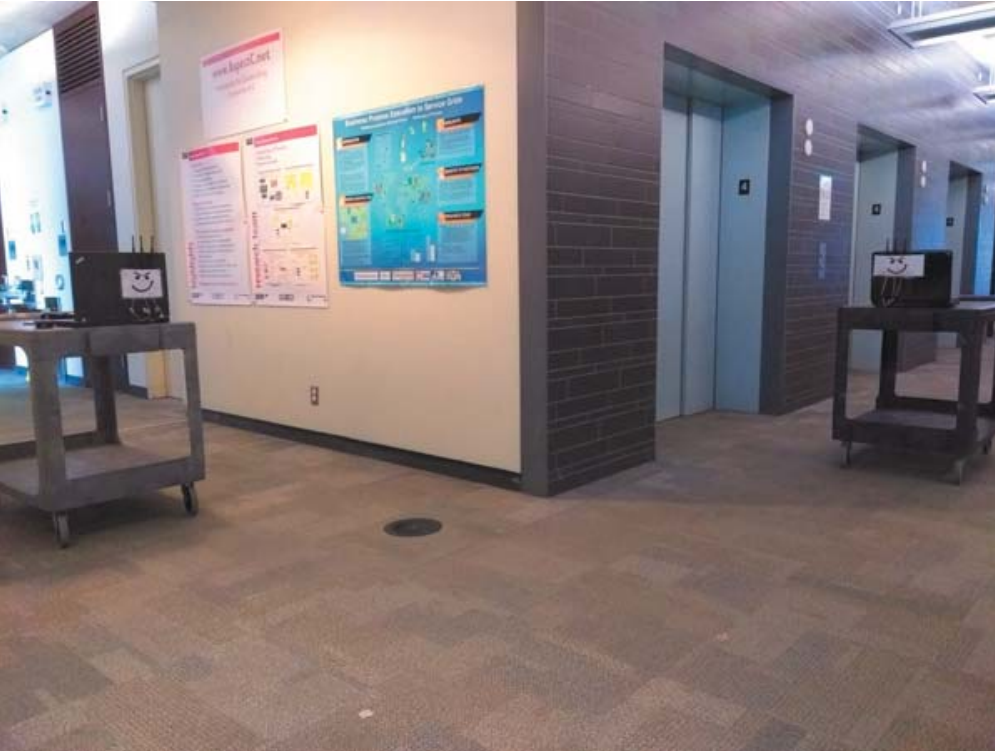}} 
\subfloat[Los (chamber).]{\label{fig12c}\includegraphics[width=3.7cm,keepaspectratio]{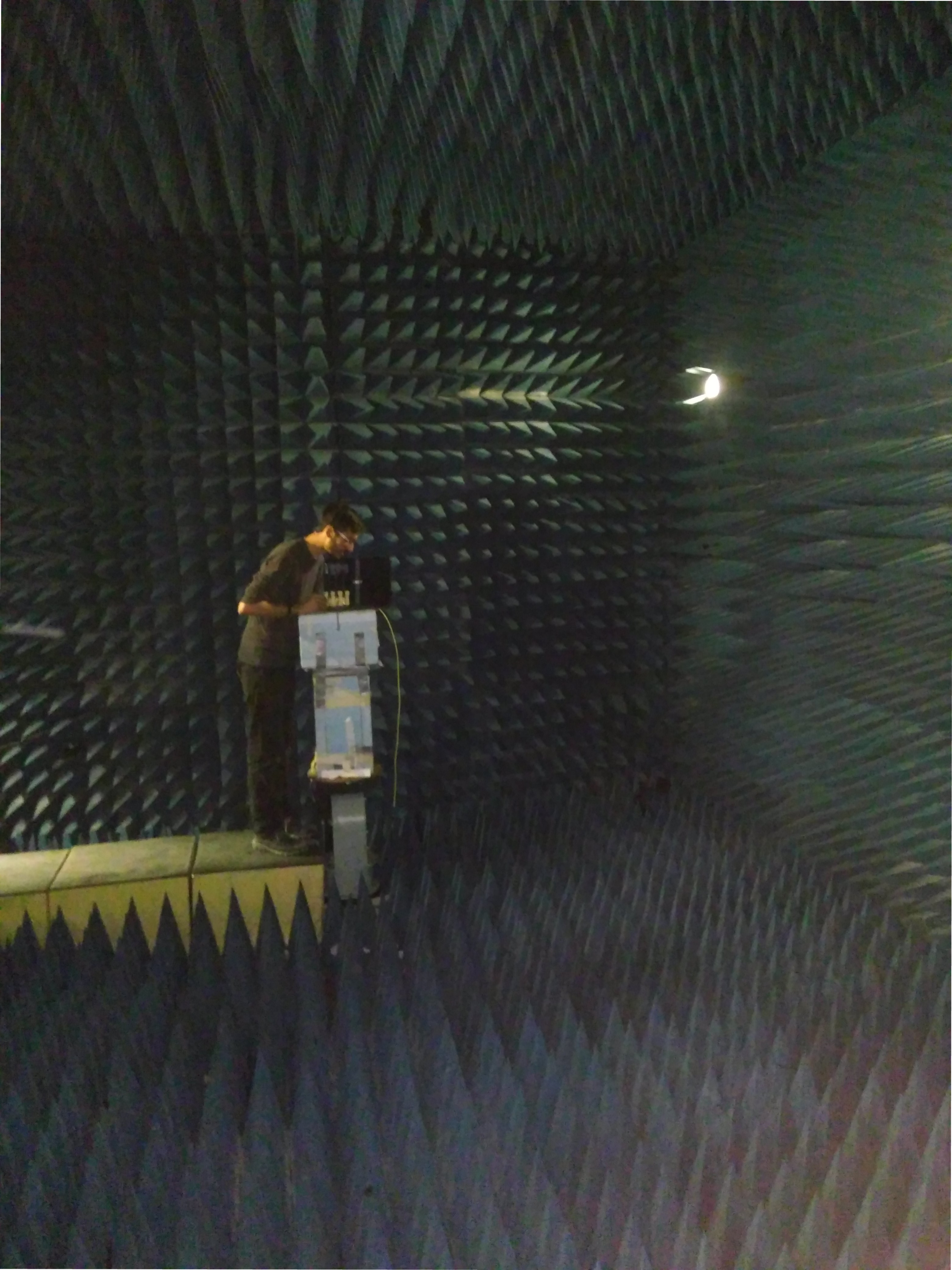}}
\subfloat[Conducted.]{\label{fig12d}\includegraphics[width=4cm,keepaspectratio]{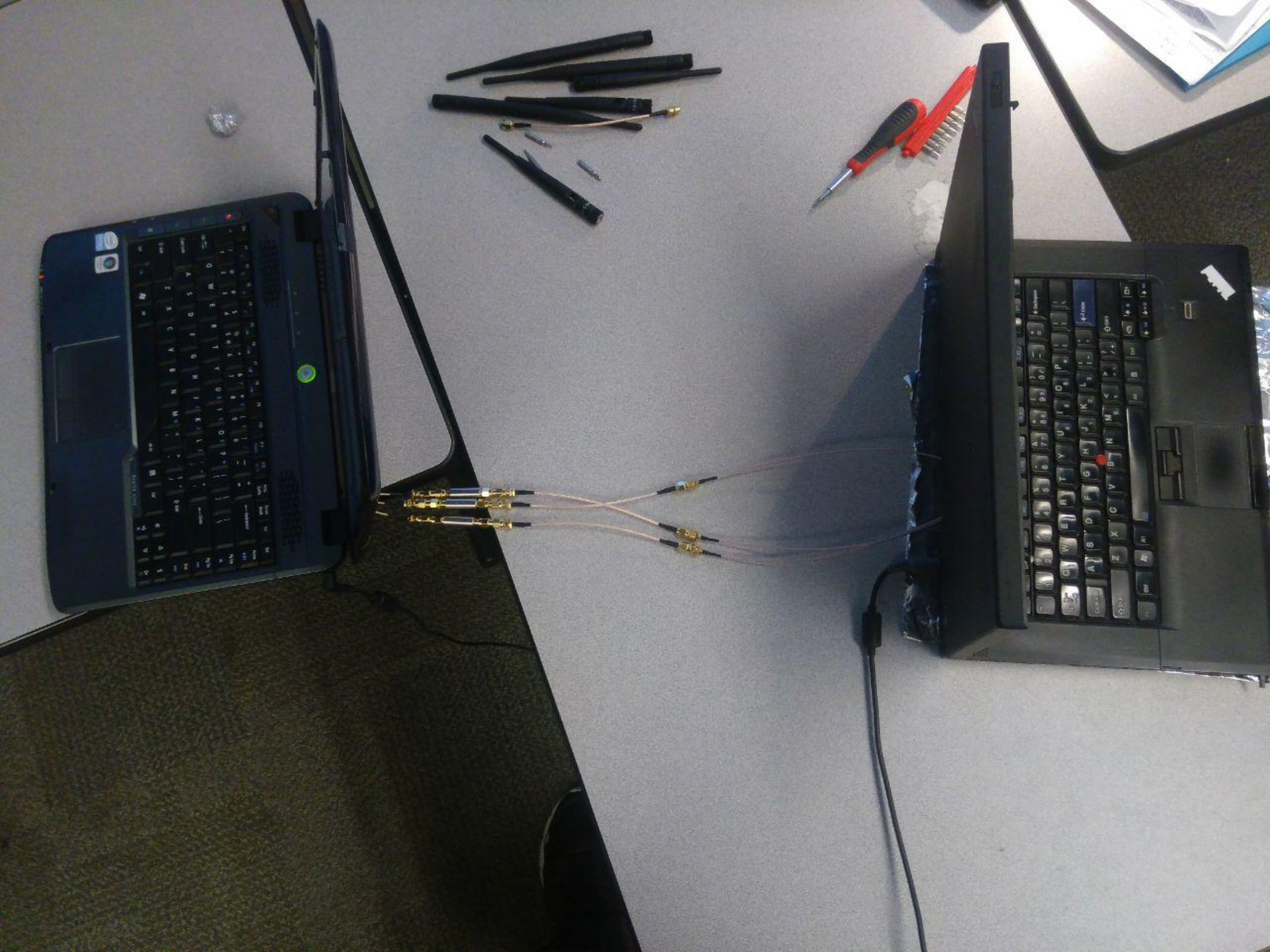}}
\caption{Different test setup investigated in this research.}
\label{fig.12}
\end{figure*}

Interestingly, $[\PsiOne_2(\timeDisc)]_{2\pi}$ is uniformly distributed $\mathcal{U}(-\pi,\pi)$ as observed through experiments and simulations.\footnote{{{Heuristically, when a random variable $X$ has a relatively high variance, the wrapped random variable $Y=[X]_{2\pi} $ behaves uniformly.}}} Provided that the wireless channel undergoes insignificant change during $\Npkt$ CSI measurements, we leverage the weak law of large numbers in Algorithm \ref{CFOremoval} to average out $\PsiOne_2(\timeDisc)$ instead of estimating it.
%
%

%
%
%
\begin{algorithm} \scriptsize
\caption{CFO $+$ Weak Stream Removal}\label{CFOremoval}
    \hspace*{\algorithmicindent} \textbf{Input:} Output $\CSIMx^{(1:\Nsc)}(1:\Npkt^{\prime})$ of Alg. \ref{STO}
    \\
    \hspace*{\algorithmicindent} \textbf{Output:}  $\breve{\CSIMx}^{(1:\Nsc)}$
\begin{algorithmic}
\Procedure{CompensationIII}{\mbox{Output $\CSIMx$ from Alg. \ref{STO}}} 
	\State ${\bf \mathsf{P}}=1$;
	\For{$\pktidx=1:\Npkt^{\prime}$}
		\State ${\bf \mathsf{P}}={\bf \mathsf{P}}\circ \CSIMx^{(\scidx)}(\pktidx)$; \Comment{Hadamard product of consecutive CSI matrices}
	\EndFor
	\State $\breve{\CSIMx}^{(\scidx)}={\bf \mathsf{P}}^{\frac{1}{\Npkt^{\prime}}}$; \Comment{Geometric averaging}
	\\  \Comment{Below: Removes weak streams for ranging}
	\State $\mathfrak{K}={\rm Norm}(\breve{\CSIMx}^{(\scidx)})$; \Comment{a single norm across $\scidx, \rxidx, \ssidx$}
	
	\For{$\rxidx=1:\Nrx$}
		\For{$\ssidx=1:\Nss$}
			\If {${\rm Norm}(\breve{\CSIVec}_{\rxidx,\ssidx}^{(\scidx)})/\mathfrak{K}<0.1$} \Comment{a single norm across $\scidx$}
				\State $\breve{\CSIVec}_{\rxidx,\ssidx}^{(\scidx)}={\rm NaN}$; 
			\EndIf
		\EndFor
	\EndFor

\EndProcedure
\end{algorithmic}
\end{algorithm}

This only leaves CPO (also known as PLL initial phase) term $\CPOoff$. When estimating range by finding the peaks of the PDP $|\CirElem_{\rxidx,\ssidx}^{(\delDisc)}|^2$, whereby $\CirElem_{\rxidx,\ssidx}^{(\delDisc)}=\mathcal{F}_{\scidx}^{-1}(\exp(-2\pi i \CPOoff)\ChanElem_{\rxidx,\ssidx}^{(\scidx)})$ is the discrete channel impulse response (CIR) on $({\rxidx,\ssidx})$th link, TOF estimation is immune to CPO since the latter gets eliminated in $|\cdot|$ operation. This is the case even when the MUSIC algorithm is used for ToF estimation \cite{Pahlavan2004} as all subspace-based methods rely on calculating the covariance matrix, which automatically eliminates phase stagnancy. 


Finally, Algorithm \ref{CFOremoval} also removes those spatial streams $(\rxidx,\ssidx)$  that are too weak (in average power sense) as those are contaminated with more noise and can potentially deteriorate ranging accuracy.
\subsection{Dealing with Pre-advancement}\label{CSIcalib:Preadvancement}
None of what was discussed so far is able to tackle pre-advancement $\Preadvancement$. The latter is neither a constant (relative to channel) to be eliminated by high-pass filtering the data, nor too variable to be averaged out. Surprisingly, our experiments show that, in both Atheros and Intel chipsets, $\Preadvancement$ \textit{changes quicker when the channel undergoes variations and it varies slowly when the channel becomes stable}. With the lack of knowledge on the dynamics of channel and the receiver, removing $\Preadvancement$ seems almost like an impossible task.

\begin{figure*}[t!]
\centering
 \includegraphics[scale=0.5]{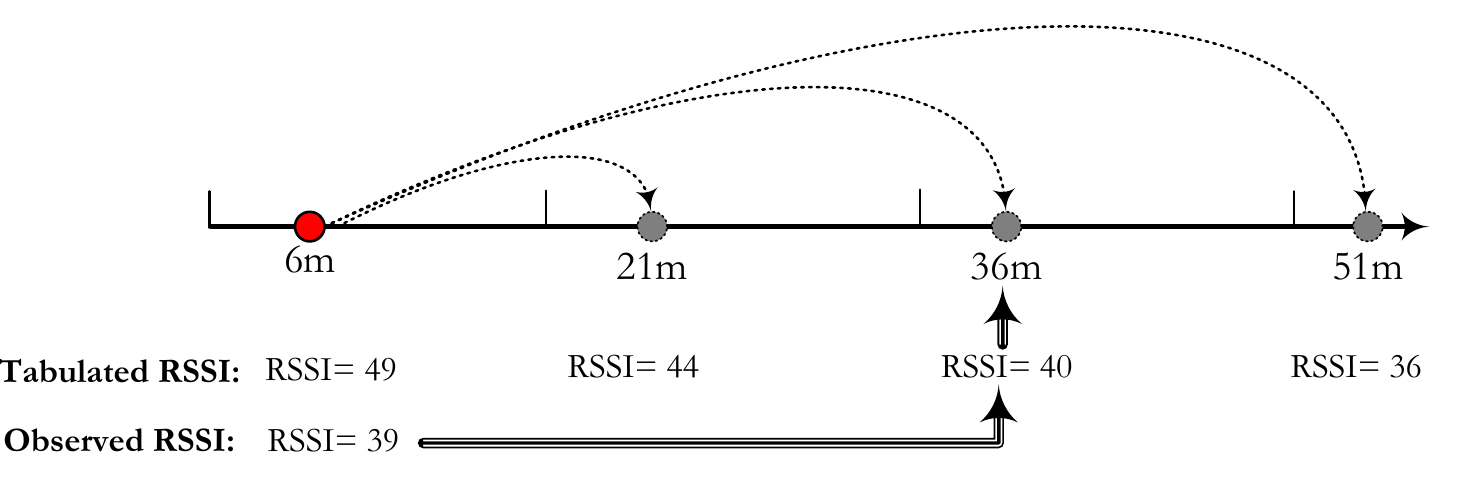}
\caption{{{Pictorial representation of RSSI assists accurate ranging estimation. Matching between the tabulated RSSI and the current packet's RSSI is used to find the true range estimate }}}
\label{fig.13}
\end{figure*}

The fact that $\Preadvancement \in \mathbb{Z}$ makes things much easier though. Let's denote the estimated transmit-receive distance obtained using post-processed CSI  after removing all contaminations by $\distEst$. Due to the presence of $\Preadvancement$, the true range $\distTruth$ is among the hypothesis set $\mathcal{D}=\{\distEst, \distEst + 15m, \distEst + 30m, \cdots\}$. Since we also have access to the received power through RSSI metric and knowing that several dB of power loss is expected as distance increases by $15$m,\footnote{The exact power reduction due to path-loss depends on many factors. Per our observation, doubling the distance results in $5-10$dB power reduction} almost all of these hypotheses in $\mathcal{D}$ are rejected except one. This idea is better illustrated in Fig. \ref{fig.13}. Briefly, all the hypotheses are formed and, then, examined based on the matching between the tabulated RSSI (collected in offline phase) and the observed RSSI (collected in online phase). In the example of Fig. \ref{fig.13}, this approach chooses $\distEst + 30=36$m as the range estimate instead of the initial $\distEst=6$m.\footnote{Note that RSSI is a relative index. Each chipset manufacturer can define their own ``RSSI-Max" value. Cisco, for example, Cisco uses a $0-100$ scale, while Atheros uses $0-60$. Nonetheless, the higher the RSSI value is, the better the signal is. In case of Atheros, RSSI=$95$+$E$(dbm), where $E$ is the received signal energy.}

\subsection{Removing Baseband Effects}\label{CSIcalib:BBeffects}
Having eliminated the synchronization sources of error, the CSI output of Algorithm \ref{CFOremoval} is given by
\begin{equation}\label{eq.21}
\breve{\CSIElem}_{\rxidx,\ssidx}^{(\scidx)}=\sum_{\txidx=1}^{\Ntx}e^{-\frac{2\pi i\scidx\big({\CDDaCoeff}_{\txidx}+{\CDDbCoeff}_{\ssidx}\big)}{\Nsc}} \ChanElem^{(\scidx)}_{\rxidx,\txidx} \SMMElem_{\txidx,\ssidx}^{(\scidx)}+\NoiseSamp_{\scidx}
\end{equation}
As discussed before, CDD affects ranging as if transmitter/receiver are farther away from each other.\footnote{For instance, with sampling time $\SampInt=50$ns  (in case of IEEE802.11n WLANs), a $d=400$ns cyclic shift is equivalent to $\delta=d/\SampInt=8$} On the other hand, SMM makes the receiver believe that there are more multipath arrivals than there really is. While it might appear that removing CDD/SMM is a trivial task, this is an implementation-dependent matter whereof details are not always available from the chipset manufacturers. As a matter of fact, the same chipset might use different CDDs at different times.  Luckily, our experiments in a conducted test setup with both Intel and Atheros chipset shows that when the channel is full rank, direct mapping takes place, where $\SMM^{(\scidx)}=\identityMx$, and CDD is always removed by the receiver. This has not been the case when $\Nss<\Ntx$. 

%
%
In conclusion, when $\Nss=\Ntx$, the CSI matrix $\CSIMx^{(\scidx)}$ is the closest to the channel matrix $\ChanMx^{(\scidx)}$ whose elements are given by 
{{\begin{equation} \label{CSIFinal}
\breve{\CSIElem}_{\rxidx,\txidx}^{(\scidx)}=\ChanElem^{(\scidx)}_{\rxidx,\txidx}=\sum_{\mpidx=1}^{\Nmp}\beta_{\mpidx}^{\rxidx,\txidx}\cdot e^{-2\pi i\freq_\scidx\DelCoeff_{\mpidx}^{\rxidx,\txidx}}+\NoiseSamp_{\scidx}
\end{equation}}}

\section{CSI-based ToF Estimation} \label{TOAestimation}

\subsection{Spectral-Domain MUSIC Algorithm}
Having cleaned the CSI samples from random and deterministic errors, our goal in this section is to use them to obtain range estimates. As discussed before, estimating the ToF of the signal using PDP has limited resolution due to bandwidth limitations of WiFi signals. On the other hand, \cite{Pahlavan2004} discovered that the subspace-based MUSIC algorithm that was traditionally used to estimate the AoA of the signal can also be used to estimate the ToF of the signal. The inherent appeal of MUSIC algorithm for ranging lies in the fact that its resolution is not only determined by the signal bandwidth but also the total signal-to-noise ratio (SNR). For that reason, MUSIC is among  \textit{super-resolution} algorithms. For more comprehensive treatment of the topic, readers are referred to \cite{Schmidt1986, Pahlavan2004, Stoica1989}. MUSIC leverages the structure of the received samples in \eqref{CSIFinal} to form
\begin{equation} \label{MUSICequation}
\breve{\CSIVec}=\SteeringMx\cdot \SourceMx+\NoiseVec
\end{equation}
where $\breve{\CSIVec}=[\breve{\CSIElem}^{(1)}\cdots\breve{\CSIElem}^{(\Nnz)}]^{\rm T}$ is a vector of frequency domain post-processed CSI samples and $\hat{\CSIVec}=\SteeringMx\cdot \SourceMx$ is the noise-free CSI vector. Based on \eqref{CSIFinal}, the steering matrix $[\SteeringMx]_{\Nnz\times \Nmp}$ , steering vector $[\SteeringVec(\DelCoeff_{\mpidx})]_{\Nnz\times 1}$, source vector $[\SourceMx]_{\Nmp\times 1}$, and  noise vector $[\NoiseVec]_{\Nnz\times 1}$ are given by
\begin{equation} \label{MUSICequation2} \small
\begin{split}
\SteeringMx= & \left[\SteeringVec(\DelCoeff_{1})\cdots \SteeringVec(\DelCoeff_{\Nmp})\right] 
\\ &
\mbox{where \;\;} \SteeringVec(\DelCoeff_{\mpidx})=\left[1, e^{-2\pi i \ScSpacing \DelCoeff_{\mpidx}}, \cdots, e^{-2\pi i (\Nnz-1)\ScSpacing \DelCoeff_{\mpidx}}  \right]^{\rm T}
\\&
\SourceMx=\left[\CompAttCoeff_1\cdots \CompAttCoeff_{\Nmp}  \right]^{\rm T}
\\ &
\CompAttCoeff_{\mpidx}=\AttCoeff_{\mpidx}e^{-2\pi i \freq_0 \DelCoeff_{\mpidx}} \mbox{\;\; and \;\;}
 \NoiseVec=\left[\NoiseSamp_1 \cdots \NoiseSamp_{\Nnz} \right]^{\rm T}
\end{split}
\end{equation}
Assuming independence of noise $\NoiseVec$ from the signal $\SourceMx$ in \eqref{MUSICequation} and $\Nnz>\Nmp$, the covariance matrix of $\breve{\CSIVec}$ is given by $\CovMx_{\breve{\CSIVec}}=\CovMx_{\hat{\CSIVec}}+\CovMx_{\NoiseVec}$ where the noise-free CSI covariance matrix $\CovMx_{\hat{\CSIVec}} =\SteeringMx \CovMx_{\SourceMx} \SteeringMx^{\rm H}$ is only of rank $\Nmp$ (rank deficient). Therefore, the largest $\Nmp$ eigenvalues in decomposition $\CovMx_{\breve{\CSIVec}}=\EigVecMx\EigValMx\EigVecMx^{\rm H}$ are due to signal (multipath arrivals) and the rest are due to noise. This observation is then used to separate the noise subspace from the signal subspace by forming the following pseudo-spectrum
\begin{equation} \label{PseudoSpectrumMUSIC}
{\rm PS}(\DelCoeff)=\dfrac{\SteeringVec(\DelCoeff)^{\rm H}\SteeringVec(\DelCoeff)}{\SteeringVec(\DelCoeff)^{\rm H}\NoiseEigMx \NoiseEigMx^{\rm H}\SteeringVec(\DelCoeff)}
\end{equation}
which ideally peaks at about $\DelCoeff=\DelCoeff_{1}\cdots \DelCoeff_{\Nmp}$ (if multipaths are sufficiently apart). In \eqref{PseudoSpectrumMUSIC}, $\NoiseEigMx=[\NoiseEigVec_{1},\cdots,\NoiseEigVec_{\Nnz-\Nmp}]$ are the noise eigenvectors corresponding to $\Nnz-\Nmp$ smallest eigenvalues of $\CovMx_{\breve{\CSIVec}}$.
For the MUSIC algorithm to work, the following conditions are to be met \cite{Stoica1989}:
\begin{enumerate}[(i)]
\item $\Nsc>\Nmp$ and $\SteeringVec(\DelCoeff_{\mpidx}) \nparallel\SteeringVec(\DelCoeff_{\mpidx^{\prime}}), \; \forall \mpidx\neq \mpidx^{\prime}$
\item $\mathbb{E}\{\NoiseVec\}=0$,  $\mathbb{E}\{\NoiseVec \NoiseVec^{*} \}=\NoisePower \identityMx$, $\mathbb{E}\{\NoiseVec \NoiseVec^{\rm T} \}=0$ (spatial whiteness)
\item $\CovMx_{\SourceMx}$ is non-singular (positive definiteness)
\end{enumerate}
It is violation of (iii) that causes the MUSIC algorithm to completely fail. The latter is indeed the case when ranging with CSI in indoor environment due to the complete coherence between source vectors $\SourceMx$ obtained for each new snapshot. Note that different snapshots are needed to calculate the empirical covariance matrix $\hat{\CovMx}_{\breve{\CSIVec}}=1/\Npkt\sum_{\pktidx=1}^{\Npkt}{\breve{\CSIVec}(\pktidx)\breve{\CSIVec}(\pktidx)^{\rm H}}$ as ${\CovMx}_{\breve{\CSIVec}}$ is never given in practice. Provided that time-domain averaging is ineffective, we perform averaging in other domains as discussed next.
\subsection{Spectral Smoothing}
Since $\Nsc\gg \Nmp$ and CSI are obtained by uniform sampling of CFR in the frequency domain, they possess invariant structure. The latter property means that the CSI vector can be partitioned into $\Npart$ spectral partitions of length $\Nsc^{\prime}$($>\Nmp$) to perform averaging across those partitions by treating them as time samples. The idea behind spectral smoothing can be explained with an example; when $\Npart=2$, \eqref{MUSICequation} is written as
\begin{equation} \small
\left[
\begin{array}{c}
\breve{\CSIVec}_{1}\\
\breve{\CSIVec_{2}}
\end{array}
\right]
=
\left[
\begin{array}{c}
\SteeringMx_{1}\\
\SteeringMx_{2}
\end{array}
\right]\cdot \SourceMx+\NoiseVec
\end{equation}
where, $\breve{\CSIVec}_{1}=\breve{\CSIVec}^{(1:\Nsc/2)}$, $\breve{\CSIVec_{2}}=\breve{\CSIVec}^{(\Nsc/2+1:\Nsc)}$, $\SteeringMx_{1}={\SteeringMx}_{(1:\Nsc/2,1:\Nmp)}$, and $\SteeringMx_{2}={\SteeringMx}_{(\Nsc/2+1:\Nsc,1:\Nmp)}$. Now, given the definition of $\SteeringVec(\DelCoeff)$ in \eqref{MUSICequation2},  $\SteeringMx_{1}= \SteeringMx_{2}\cdot{\bf M}$ where ${\bf M} ={\rm diag}(z_1^{\Nsc/2}\cdots z_{\Nmp}^{\Nsc/2}), \;\; z_{\mpidx}=\exp(-2\pi i \ScSpacing \DelCoeff_{\mpidx})$.

With this property, a hardened covariance matrix can be obtained by averaging individual sub-array's covariance matrices as $\CovMx_{\rm ss}=0.5(\CovMx_{\breve{\CSIVec}_{1}}+\CovMx_{\breve{\CSIVec}_{2}})=\SteeringMx_{1}\CovMx_{\SourceMx}^{\prime}\SteeringMx_{1}^{\rm H}$ where $\CovMx_{\SourceMx}^{\prime}=0.5(\SourceMx\SourceMx^{\rm H}+{\bf M}\SourceMx\SourceMx^{\rm H}{\bf M}^{\rm H})$ has an improved rank, thus closer to the true source covariance matrix. In the general case, covariance hardening can be achieved if $\Npart\geq \Nmp$ \cite{Kailath1985} through:
\begin{equation}\label{CovMxFw}
\begin{split}
\CovMx_{\rm ss}= 
&
\frac{1}{\Npart}\sum_{\partidx=1}^{\Nsc-\Nsc^{\prime}+1}{\CovMx_{\breve{\CSIVec}_{\partidx}}^{(\partidx:\partidx+\Nsc^{\prime}-1,\partidx:\partidx+\Nsc^{\prime}-1)}} 
\\ &
\mbox{s.t .\;\;} (\partidx+\Nsc^{\prime}-1 \leq \Nsc/2 \mbox{{ \bf or }} \partidx\geq \Nsc/2+1)
\end{split}
\end{equation}
where the constraints above are to ascertain that no sub-array contains $\scidx=0$ subcarrier. This is an important consideration as no information is sent on $\scidx=0$ (due to large DC current) causing $\CSIElem^{(\scidx=-1)}$ and $\CSIElem^{(\scidx=1)}$ to be $2 \ScSpacing$ apart instead of $\ScSpacing$ (as is the case for other neighbouring subcarriers). Not paying attention to this  when performing spectral smoothing adds to the estimation error.

Since the averaged covariance matrix has the exact structure as the original covariance matrix, MUSIC can be applied to obtain ToF estimates. This advantage came at the cost of reducing the effective length of the CSI vector ($\Nsc \rightarrow \Nsc^{\prime}$) and results in a tradeoff between resolution and stability of MUSIC solution.


%

\subsection{Forward-Backward Smoothing}
The invariant structure of the CSI signal model can be used not only to smooth over the forward covariance sub-matrices but also the backward ones \cite{Kwon1989}. Mathematically, this operation is equivalent to
\begin{equation}\label{CovMxFwBw}
\CovMx_{\breve{\CSIVec}}^{\rm FW/BW}=\CovMx_{\breve{\CSIVec}}^{\rm FW}+\underbrace{{\antidiagMx}{\CovMx_{\breve{\CSIVec}}^{\rm FW}}^{*}{\antidiagMx}}_{\CovMx_{\breve{\CSIVec}}^{\rm BW}}
\end{equation}
where $\antidiagMx$ is an $\Nsc \times \Nsc$ exchange matrix with ones only on the anti-diagonal and $\CovMx_{\breve{\CSIVec}}^{\rm FW}$ is the forward covariance matrix which is given by \eqref{CovMxFw}. Using forward-backward smoothing, the empirical covariance matrix in \eqref{CovMxFwBw} gets closer to the true CSI covariance matrix improving the accuracy of the MUSIC estimator that heavily relies on the knowledge of this matrix.
\subsection{Decision Fusion}
\begin{figure}[t]
\centering
\subfloat[Pseudo-spectrums.]{\label{fig10a}\includegraphics[scale=0.4]{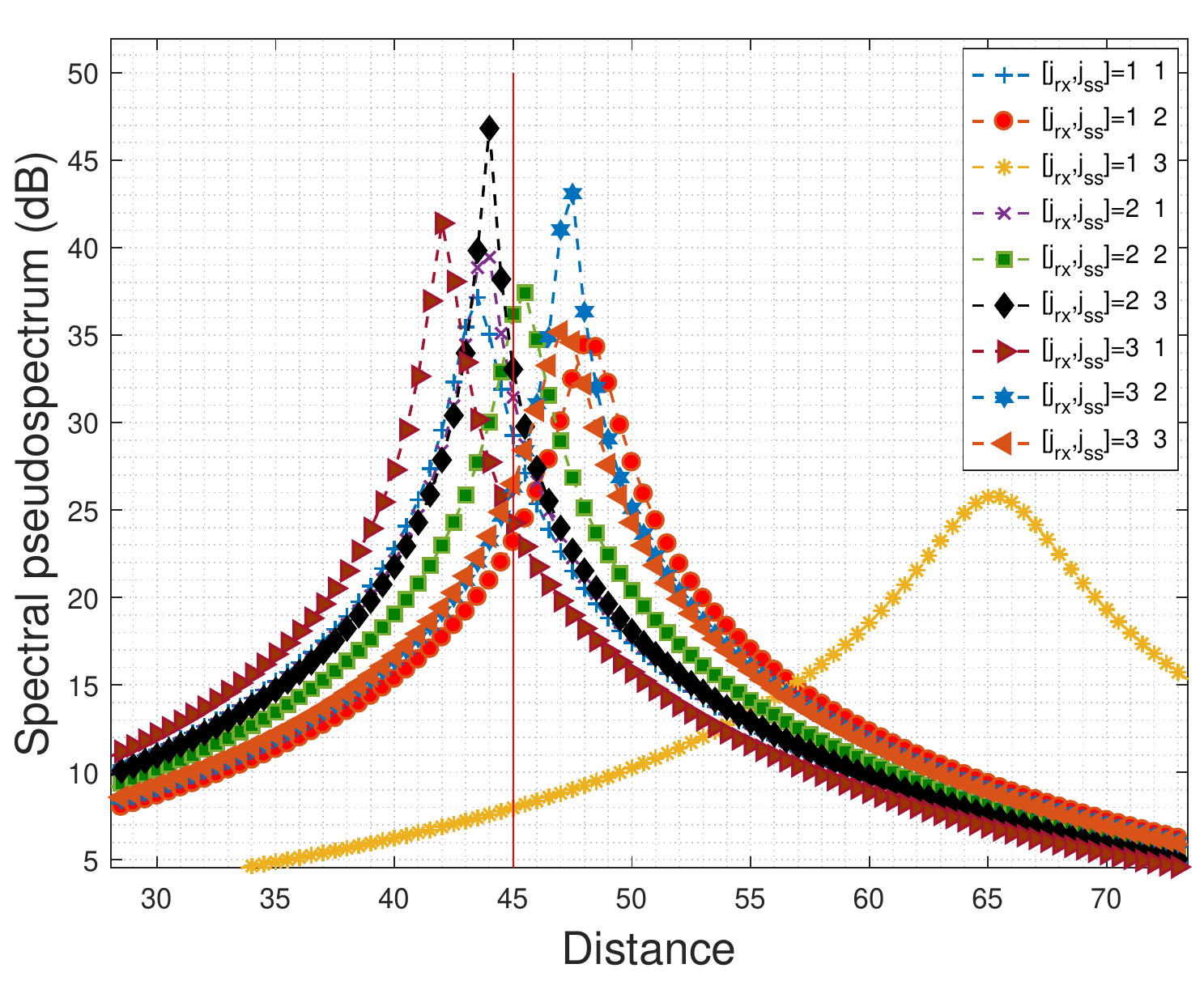}} \\
\subfloat[ECDF of estimation error.]{\label{fig10b}\includegraphics[scale=0.4]{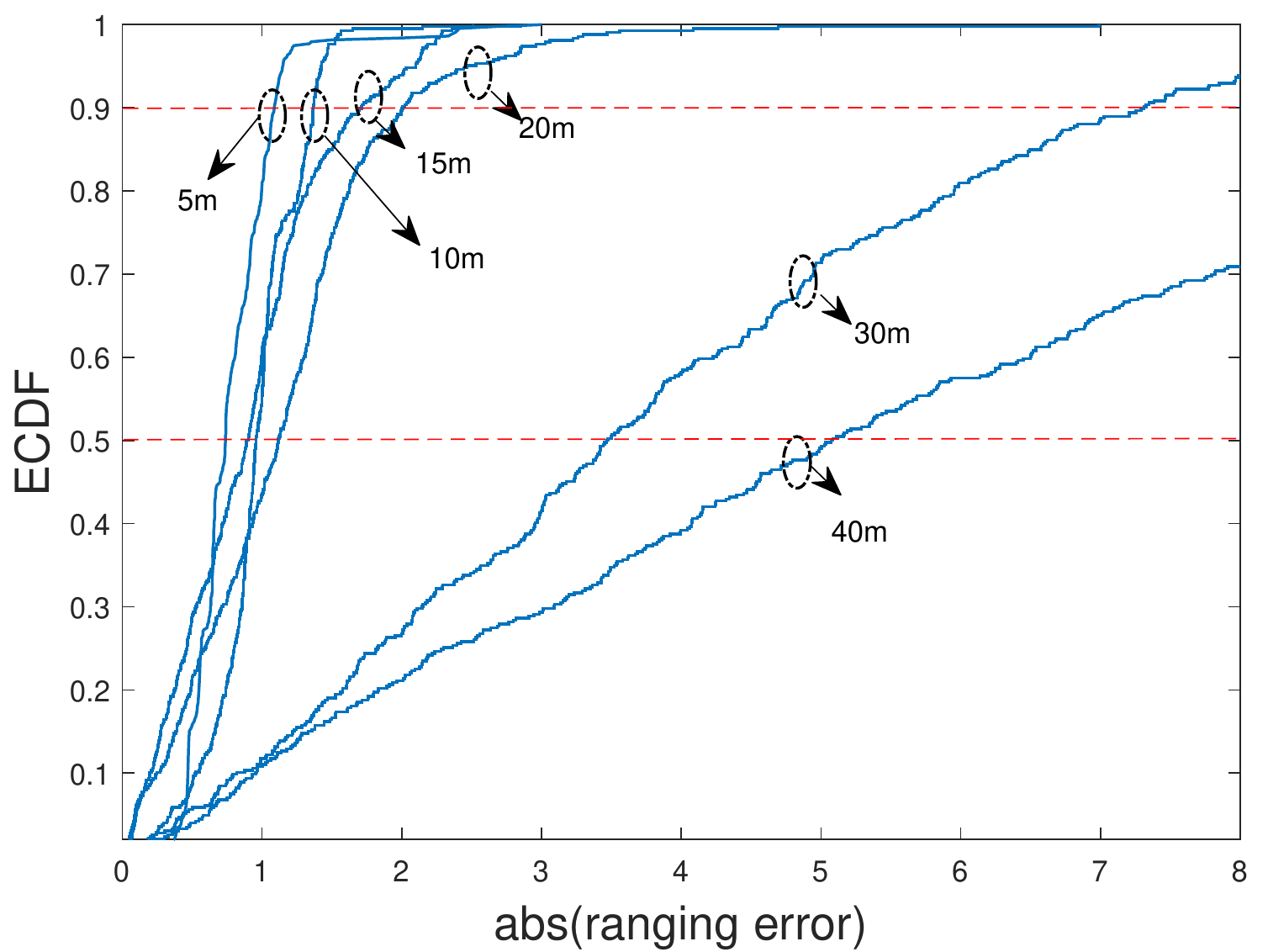}}
\caption{{{(left) Pseudo-spectrums derived for real-world measurements for different $(\rxidx,\txidx)$ pairs when transmitter-receiver are 45m apart (red vertical line). (right) ECDF of estimation error when transmitter and receiver are 5, 10, 15, 20, 30, 40m apart.}}}
\label{figPSs}
\end{figure}

As detailed before, CSI is a matrix characterizing the channels between each transmit antenna $\txidx$ and receive antenna $\rxidx$. Provided that transmit antennas (and receive antennas) are spatially sufficiently separated, CSI provides us with $\rxidx\times \txidx$ independent sub-channels which can be used to obtain independent estimates of range. By fusing these decisions at last, one expects a more accurate final range estimate. One should note that CSI over different sub-channels cannot be stacked up to form a single covariance matrix (named fusion over raw measurement), as the steering matrix $\SteeringMx$ in \eqref{MUSICequation} is different for different transmit-receive sub-channels. Fig. \ref{figPSs} illustrates pseudo-spectrums calculated from measurement in corridors of University of Toronto (UofT) where transmitter and receiver were 45m apart.

%
%
%

%
Different pseudo-spectrums peak at different ranges ($\distEst_{\rxidx,\txidx}$) for two reasons: \textbf{(i)} some $(\rxidx,\txidx)$ sub-channels are weaker thus giving rise to larger random shifts in their corresponding pseudo-spectrums \textbf{(ii)} multipath fading impacts sub-channels differently. 

How do we combine $\distEst_{\rxidx,\txidx}$s to get a more accurate range estimate? Simply averaging them yields inaccurate range estimates. The solution we propose is  to obtain all sub-channel pseudo-spectrums for all post-processed CSI packets and:
\begin{enumerate}
\item identify those $\distEst_{\rxidx,\txidx}$ that fluctuate the most (in time). Apply any outlier rejection method to remove that sub-channel from decision-making process. For example, that peak is $\distEst_{1,3}$ in Fig. \ref{figPSs}.
\item weight the remaining pseudo-spectrum peaks with their corresponding averaged (w.r.t $\scidx$) CSI magnitudes, i.e.  $w_{\rxidx,\txidx}=1/\Nsc\sum_{\scidx}{|\CSIElem^{(\scidx)}_{\rxidx,\txidx}|}$. The logic is that a weaker sub-channel is more impacted by the noise components when CSI matrix is being estimated by the receiver through simple zero-forcing (ZF) or minimum-mean-square (MMSE) algorithms. Since the weakness/strength of a sub-channel is manifested in the total energy spread across frequency components, the above weighting put more emphasis on the stronger sub-channels that are less impacted by estimation noise.
\end{enumerate}

\section{Experimental Results}\label{ExpResults}
We performed extensive experiments using IEEE802.11n Atheros 93xx chipset in two environments: (i) Anechoic chambers (multipath-free) in Fig. \ref{fig12c} (ii) corridors (multipath) in Fig. \ref{fig12a} and Fig. \ref{fig12b}. We collected a few thousands  CSI, post-processed them using the techniques introduced in Section \ref{CSIcalib}, formed the spatially-smoothed covariance matrix, applied spectral MUSIC algorithm, fused final decisions, and obtained one final estimate. We repeated this experiment for all the collected CSI set. Fig. \ref{fig10b} plots the empirical cumulative distribution function (ECDF) of the estimation error: {{According to this plot, the median accuracy of 60, 80, 90, 115, 140, 350, and 500cm}} is achieved when transmitter-receiver are 5, 10, 15, 20, 25, 30, and 40m apart respectively. The 90th percentile accuracy is about 1m (at 5m) and 1.7m (at 20m). Note that at 40m distance, the received power is about -100dB which is only a few decibels higher than the noise floor. {{One should note that using raw CSI yields range estimates that are off by several tens of meters. Provided that the range estimation error is even larger than the maximum WiFi coverage, which is about 20m in current MIMO-OFDM systems, there is truly no value in comparing calibrated and uncalibrated range estimates. Moreover, since our work is the first to exploit the CSI phase to obtain range estimates, it was impossible to benchmark our results with the previous studies that are based on angle of arrival (AoA) estimation.}}

Contrary to the claims made in  the literature \cite{kotaru2015, vasisht2016}, our results suggest that sub-meter ranging accuracy is possible using CSI obtainable from commodity WiFi. 
%
%
There are several observations that were made in the course of the project which are worth mentioning:
\begin{itemize}
\item   With $20$MHz of spectrum available for WiFi signals, one shall not expect to resolve multipath components with MUSIC or with any high-resolution estimation algorithm. This is evident from Fig. \ref{fig10a} for 9 sub-channels of a highly fading propagation environment. 
\item Due to the limited resolvability power achievable with $20$MHz CSI, spectral smoothing with only a handful of partitions is able to sufficiently harden the covariance matrix and recover the only expected peak.
\item When SNR is low (which happens at longer distances), receiver chooses to advance more than one symbol to make sure any mistake in detecting the symbol boundary estimation (using cross-correlation of HT-LTF sequence with the received header) wouldn't cause erroneous outcomes.
\item In situations when $\distTruth<\Preadvancement$, the pseudo-spectrum has a (one-sided) peak at $d=0$ which signifies the existence of a hidden (two-sided) peak at negative distances. When that's the case, the RSSI-assisted approach won't work as the knowledge of the hypothesis set $\mathcal{D}=\{\distEst, \distEst + 15m, \distEst + 30m, \cdots\}$, delineated in sub-section \ref{CSIcalib:Preadvancement}, hinges on the knowledge of the two-sided $\distEst$. To cope with this situation, after post-processing CSI, one will have to deliberately rotate (leftwise) the CSI phase (before forming the pseudo-spectrum) by a few samples to recover peaks at negative distances (due to the existence of pre-advancement error) and de-rotate those peaks for the \textit{same number of samples} to cancel out what was artificially added. Then RSSI-based hypothesis testing is applied to find out the true range estimate.
\item This work does not consider tracking of user's range parameter through combining motion information (obtainable using prevalent inertial measurement units (IMU)) with instantaneously obtained range estimates. Therefore, it is believed that such fusion of information (e.g. using \textit{Kalman} filter) would yield more stable and accurate results.
\end{itemize}


%
%

\section{Conclusion}
The availability of channel-state information (CSI) from WiFi chipsets has made indoor positioning a reality. Leveraging the CSI, several recent studies have achieved decimeter accuracy through angle-of-arrival (AoA) estimation or wideband ranging. When it comes down to implementation, the CSI-based localization with time-of-flight (ToF) measurement from only a single channel ($20$ MHz of spectrum) has either not been pursued or led to inconsistent results. With the knowledge of fundamental limits of ranging and its reliance on the availability of bandwidth, the critical question has always been ``whether sub-meter ranging is possible with such limited bandwidth". This paper aims to answer this question. We dissect different deterministic and random phenomena happening in the transmitter and receiver hardware and establish the right model for CSI. We propose techniques to eliminate random phases introduced by the insufficiency of synchronization between transmitter and receiver. Our range estimates using the MUSIC algorithm show that median accuracy of $0.6$m ($1.15$m) is achievable in highly multipath line-of-sight environment where transmitter and receiver are $5$m ($20$m) apart. Moreover, with $90$th percentile accuracy of $1.1$m ($2$m) in $5$m ($20$m), we can claim that the proposed system is robust.
%

\end{document}